\documentclass{aa}
\usepackage{graphicx}
\def\ee{\end{equation}}
\def\be{\begin{equation}}
\def\bdm{\begin{displaymath}}
\def\edm{\end{displaymath}}

\usepackage{graphics}
\begin{document}
\def\be{\begin{equation}}
\def\ee{\end{equation}}
\def\bdm{\begin{displaymath}}
\def\edm{\end{displaymath}}
\def\sk{\rm {sgn}(k)}
\def\sq{\rm {sgn}(q_i)}
\def\sN{\rm {sgn}(N)}

\title{Conversion of relativistic pair energy into radiation in the jets
of active galactic nuclei}
\author{R. Schlickeiser \inst{1}, 
R. Vainio \inst{2}, 
M. B\"ottcher \inst{3,4,6},  
I. Lerche \inst{5}, 
M. Pohl \inst{1}
\and C. Schuster \inst{1}}
\institute{
$^1$ Institut f\"ur Theoretische Physik, Lehrstuhl IV: 
Weltraum- und Astrophysik, Ruhr-Universit\"at Bochum, 
D-44780 Bochum, Germany\\
$2$ Space Research Laboratory, Department of Physics, 
Turku University, FIN-20014 Turku, Finland\\
$3$ Department of Physics and Astronomy, Rice University, MS 108, 
6100 S. Main Street, Houston, TX 77005 - 1892, USA\\
$4$ Department of Physics and Astronomy, Ohio University, Clippinger 339, Athens,
OH 45701, USA (current address)\\
$^5$ Department of Geological Sciences, University of South 
Carolina, Columbia, SC 29208, USA\\
$^6$ Chandra Fellow}

\date{Received October 30, 2001; accepted June 28, 2002}

\offprints{R. Schlickeiser; rsch@tp4.ruhr-uni-bochum.de}
%\date{ }
\authorrunning{Schlickeiser et al.}
\titlerunning{Conversion of relativistic pair energy}

\abstract{It is generally accepted that relativistic jet outflows power the 
nonthermal emission from active galactic nuclei (AGN). The
composition of these jets -- leptonic versus hadronic -- is still under debate. 
We investigate the
microphysical details of the conversion process of the kinetic energy 
in collimated relativistic pair outflows into radiation through interactions 
with the ambient interstellar medium. 
Viewed from the coordinate system
comoving with the pair outflow, the interstellar protons and electrons 
represent a
proton-electron beam propagating with relativistic speed in the pair plasma.  
 We demonstrate that the beam 
excites both electrostatic and 
low-frequency magnetohydrodynamic Alfven-type waves via a two-stream
instability in the pair background plasma, and we calculate the time evolution 
of the distribution functions of the
beam particles and the generated
plasma wave turbulence power spectra.
For standard AGN jet outflow and environment parameters we show that
the initial 
beam distributions of interstellar protons and electrons quickly relax
to  plateau-distributions 
in parallel momentum, transferring thereby one-half 
of the initial energy density of the beam particles to electric field
fluctuations of the generated electrostatic turbulence.
On considerably longer time scales, 
the plateaued  interstellar electrons and protons will isotropise by their self-generated
transverse turbulence and thus be picked-up in the outflow pair plasma.
These longer time scales are also characteristic for the development of
transverse hydromagnetic turbulence 
from the plateaued electrons and protons. This hydromagnetic 
turbulence upstream and downstream is crucial for diffusive shock acceleration
to operate at external or internal shocks associated with pair outflows.   
}

 \maketitle
%
%________________________________________________________________

\section{Introduction}
The detection of intense medium-energy gamma radiation from over 60 blazar 
active gactic nuclei (hereafter abbreviated as AGNs) 
with the EGRET instrument on the Compton observatory
(Hartman et al. \cite{h99}) and TeV gamma radiation from several BL-Lac AGNs (Pohl 
\cite{p01})
shows that nonthermal gamma-ray production is a significant dissipation mechanism
of jet energy generated by black-hole accretion. Besides the modelling of the 
broadband nonthermal radiation, 
the composition of the jet plasma -- i.e.  
electron-positron pair jets (leptonic jets) versus electron-proton jets 
(hadronic jets) -- and the acceleration of these jet particles to relativistic 
energies are main subjects of the current theoretical research. 

Gamma radiation in leptonic
models of broadband blazar emission is attributed to synchrotron self-Compton 
(Maraschi et al. \cite{mgc92}; Bloom \& Marscher \cite{bm96};
Tavecchio et al. \cite{tmg98}) or external Compton (Dermer \& Schlickeiser \cite{ds93}; 
Sikora et al. \cite{sbr94}; B\"ottcher et al. \cite{bms97}; Dermer et al. \cite{dss97}; 
Arbeiter et al. \cite{aps02}) processes (see B\"ottcher \cite{b01} and 
Sikora \& Madejski \cite{sm01} for recent reviews). In hadronic models, secondary 
photopairs and photomesons and secondary mesons are produced when energetic protons and
ions interact either with ambient synchrotron photons (Mannheim \& Biermann 
\cite{mb92}; Mannheim \cite{m93}), photons of the external field (Bednarek \& 
Protheroe \cite{bp99}; Atoyan \& Dermer \cite{ad01}) and/or ambient matter fields 
(Pohl \& Schlickeiser \cite{ps00}). Observationally, future detection 
of high-energy neutrino emission correlated with high-energy photon emission 
(Schuster et al. \cite{sps02}) will provide the ultimate 
test between the leptonic and hadronic jet models.

Most existing radiation models of AGN jets are very unspecific on the microphysical 
details of the conversion of the kinetic jet energy into 
energetic charged particles and subsequently into radiation.
Without detailed discussion these models  
often assume 
that a significant fraction of the accreted kinetic energy is injected into 
nonthermal pairs and/or hadrons with power-law distribution functions. This efficient
conversion is attributed to the scenario that the outflowing relativistic jet plasma 
has produced a relativistic shock with fully developed
hydromagnetic turbulence in order to allow for efficient non-thermal diffusive 
particle acceleration at the collisionless shock fronts:

\noindent (1) This neglects the fact that it takes a
finite time to build up the necessary turbulence in the two-stream
multi-fluid system consisting of the relativistically moving jet plasma and
the traversed interstellar or intergalactic hydrogen plasma. 

\noindent (2) It is not
clear from the beginning that the outflowing relativistic jet plasma will 
generate primarily (and enough) transverse magnetohydrodynamic turbulence, 
which is crucial for efficient particle deflections in the up- and downstream 
region of the shock waves. It is well conceivable that most of the kinetic blast
wave energy is transferred to electrostatic plasma turbulence and not to
transverse hydromagnetic turbulence.

\noindent (3) The properties of the generated plasma
turbulence are decisive both for the formation and the nature of 
the developing collisionless shock waves. It is known from 
non-relativistic shock theory that the inclusion of the finite pressure and 
energy flux of the generated plasma turbulence in the Rankine-Hugoniot shock
relations strongly 
modifies the standard fluid shock properties (Vainio \& Schlickeiser 
\cite{vs99, vs01}; Lerche et al. \cite{lps00}) and subsequently 
the energy spectrum of the accelerated nonthermal particles.

It is the purpose of this work to consider more thoroughly some of the
microphysical details of the energy conversion in relativistic jet outflows.
We consider the energisation of relativistic particles in the jet  
by interactions with the surrounding medium following the earlier work of three
of us (Pohl \& Schlickeiser \cite{ps00}; 
Pohl et al. \cite{pls02}). There the AGN jet has been
assumed to be a cloud of dense {\it electron-proton} plasma which moves 
relativistically
through the electron-proton interstellar medium of the AGN host galaxy. 
The plasma cloud is assumed to have a cylindrical shape with thickness
$d^*=10^{13}d^*_{13}$ cm, which is small compared to the radius
$r^*=10^{14}r^*_{14}$ cm of the cylinder (see Fig. 1 of 
Pohl \& Schlickeiser (\cite{ps00}) for a sketch of the assumed cloud geometry).
It was 
shown that in such {\it hadronic jets} swept-up
ambient matter is quickly isotropised in the jet cloud frame by a relativistic 
two-stream instability, which provides relativistic particles in the jet cloud 
without invoking any acceleration process. Here and in the following the index $*$ indicates 
the values of physical quantities 
in the laboratory (AGN) frame; quantities not indexed are in the jet frame.

Here we study the {\it leptonic variant} of this jet outflow model, i.e. we 
model the outflowing jet cloud 
as a one-dimensional channeled outflow of thickness $d^*$ with relativistic 
bulk velocity $\vec{V}$, consisting of {\it pairs of electrons and positrons} of 
density $n_b^*$ instead of {\it electrons and protons}. To avoid 
dramatic pair annihilation, the density of pairs must be limited, and we allow 
for a thermal pair distribution with
non-relativistic temperature $\Theta =k_BT_{\rm pair}/(m_ec^2)<<1$ in the 
jet rest frame. This beam of pairs 
propagates into the surrounding interstellar medium that consists
of cold protons and electrons at rest of density $n_i^*$. 

For mathematical convenience we assume that the outflow
is directed parallel to a uniform background magnetic field. The 
assumption of the magnetic field directed along the ejecta velocity enormously
facilitates the analytical treatment of the two-stream instability, but 
crucially depends on the location of the particle energisation 
with respect to the large-scale magnetic field and the confinement of the jet ejecta. 
In magnetohydrodynamic models of jets in accreting systems the poloidal 
magnetic field component is more strongly 
($\propto R^{-2}$)
suppressed by the side expansion of the ejecta than the transverse magnetic field 
component ($\propto R^{-1}$), so that our assumption will hold at close distances 
to the central object. Moreover, in the presence of a dominating transverse magnetic
field it can be argued that the ambient medium  will penetrate the ejecta by one 
Larmor radius only and the momentum exchange should be faster and more efficient 
than in the case of a dominating poloidal magnetic field considered here.
{\footnote {We are grateful to the referee for noting this difference.}}

Viewed from the coordinate system
comoving with the pair outflow, the interstellar protons and electrons 
represent a
proton-electron beam propagating with relativistic speed $-\vec{V}$ 
antiparallel to the
uniform magnetic field direction. Modifying the analysis of Pohl \& 
Schlickeiser (\cite{ps00})
and Pohl et al. (\cite{pls02}) for cold electron-proton
outflows to finite temperature pair outflows, we demonstrate that the beam 
excites both electrostatic and 
low-frequency magnetohydrodynamic Alfven-type waves via a two-stream
instability in the pair background plasma. We study the time evolution of the 
beam particles, the generated
plasma wave turbulence power spectra and discuss the radiation signatures of
such systems.
\section{Basic equations}
\subsection{Outflow parameters}
In the comoving frame 
the total phase space distribution
function of the plasma in the blast wave region at the start (time $t=0$) is
\bdm
f({p},t=0)= n_if_i(p,\mu ,t=0)+\; n_bf_b(p,\mu ,t=0)=
\edm
\be
{1\over 2\pi p^2}n_i\delta \bigl(\mu +1\bigr)\delta \bigl(p-P\bigr) +\; 
n_bT(p,\Theta ) 
\ ,\label{dist_f1}
\ee 
with the thermal pair distribution 
\be
T(p,\Theta )=\bigl[4\pi (m_ec)^3\Theta K_2(\Theta ^{-1})\bigr]^{-1}
\exp \bigl(-{\sqrt{1+{p^2\over m_e^2c^2}}\over \Theta }\bigr)
\label{pairmax}
\ee
and the dimensionless pair temperature 
\be
\Theta \equiv {k_BT_{\rm pair}\over m_ec^2}
\label{pairtempe}
\ee
Here $\mu =p_{\parallel }/p$ is
the cosine of the pitch-angle of the particles in the magnetic field
$B$, and $P=\Gamma mV=mc\sqrt{\Gamma ^2-1}$ where $\Gamma =1/\sqrt{1-(V^2/c^2)}$.
The number densities transform as $n_i=\Gamma n_i^*,\;\;\, n_b=n_b^*/\Gamma $.

Earlier radiation modelling of AGN jets (e.g. Dermer \& Schlickeiser \cite{ds93};
Pohl \& Schlickeiser \cite{ps00}) indicated that 
the following parameters are appropriate: the
density of pairs in the jet cloud is of
order $n^*_b=10^{10}n^*_{b,10}$ cm$^{-3}$, 
implying $n_b=10^8n_{b,8}=10^8n^*_{b,10}\Gamma _2^{-1}$ cm $^{-3}$, and 
much larger than 
the interstellar gas density $n^*_i$. The initial Lorentz factor is of 
order $\Gamma =10^2\Gamma _2$ and the total number of pairs in the plasmoid is 
$N_{\rm tot}=3\cdot 10^{51}(r^*_{14})^2d^*_{13}n^*_{b,10}$, corresponding 
to a total kinetic energy of 
$E_{\rm tot}=5\cdot 10^{47} \Gamma _2(r^*_{14})^2d^*_{13}n^*_{b,10}$ erg. 

The background magnetic field in the blast wave 
is parametrized by the equipartition parameter $\epsilon _B=0.1\epsilon _{B,-1}$
through
\be
B=0.14(\epsilon _{B,-1}n_{b,8}\Theta _{-4})^{1/2}
\hbox {  Gauss}
\label{bequi}
\ee
for a non-relativistic initial normalised (co-moving) pair temperature
$\Theta =10^{-4}\Theta _{-4}$. 
The plasma parameter of the pair plasma 
\bdm
g_p=
7.3\cdot 10^{-4}n_b^{1/2}[{m_ec^2\Theta \over k_B}]^{-3/2}=
\edm
\be
1.6\cdot 10^{-8}(n^*_{b,10})^{1/2}\Gamma _2^{-1/2}\Theta _{-4}^{-3/2}<<1
\label{plasmapairparameter}
\ee
is much smaller than unity, so that the kinetic plasma description is appropriate.

The beam distribution function $f_i$ in Eq. (\ref{dist_f1}) is 
unstable and excites electrostatic and low-frequency transverse plasma waves.
We want to calculate the time scales $t_e$ and $t_t$ it takes these 
plasma waves to relax the incoming
interstellar proton-electron beam to a stable distribution 
with respect to both instabilities. Of particular interest is 
the timescale $t_t$ for beam isotropisation: if this relaxation time is much
smaller than the light travel time $d/c$ then an isotropic distribution of 
protons and
electrons in the blast wave region is efficiently generated. 
\subsection{Parallel plasma waves in hot pair plasmas}
We restrict our
analysis here to parallel propagating electrostatic and low-frequency 
transverse plasma waves.
Because
\be
n_i/n_b=\Gamma ^2n^*_i/n^*_b=
10^{-6}\Gamma _2^2n_i^*/n^*_{b,10}<<1\ ,
\ee
the beam is weak. Therefore the contributions from the beam to 
the plasma wave dispersion relations of electrostatic waves and transverse waves
at frequencies $\omega _R$ well below the
non-relativistic electron gyrofrequency ($|\omega _R|<<|\Omega _e|$) are 
perturbations to the longitudinal ($\Re \Lambda _e=0$) and 
transverse dispersion relations ($\Re \Lambda _t=0$) in the thermal pair plasma.
The properties of the longitudinal and transverse dispersion relations in
a thermal pair plasma are derived in Appendix A.

From the real part of the longitudinal 
dispersion relation in the thermal pair plasma we obtain the 
electrostatic waves
\be
\omega _R^2=2\omega ^2_{p,e}
\label{epos1}
\ee
in the wavenumber range $|k|\le 2\omega _{p,e}/(c\Theta ^{1/2})$
as the only longitudinal plasma mode. These waves are not damped ($\omega _I=0$) 
in the superluminal wavenumber range $|k|<\sqrt{2}\omega _{p,e}/c$, whereas in the
subluminal wavenumber range 
$\sqrt{2}\omega _{p,e}/c\le |k|\le 2\omega _{p,e}/(c\Theta ^{1/2})$
the damping rate is
\bdm
\omega _I=
-\pi \rm {sgn}(\omega _R) 
{\omega _{p,e}^4\over ck[c^2k^2-2\omega _{p,e}^2]\Theta K_2(1/\Theta )}
\edm
\bdm
\times \exp \Bigl[-{c|k|\over \Theta \sqrt{c^2k^2-2\omega _{p,e}^2}}\Bigr]
\edm
\be
\times \Bigl[1+\; {2\Theta \over c|k|}\sqrt{c^2k^2-2\omega _{þ,e}^2}+\; 
{2\Theta (c^2k^2-2\omega _{p,e}^2)\over c^2k^2}\Bigr]
\label{damprateelect1}
\ee
where
\be
\omega _{p,e}=\sqrt{4\pi e^2n_b/m_e}
=5.64\cdot 10^8(n^*_{b,10})^{1/2}\Gamma _2^{-1/2}\hbox {   Hz}
\label{defelectronplasma}
\ee
is the electron plasma frequency. 

Likewise, from the real part of the transverse 
dispersion relation in the thermal pair plasma we obtain the 
transverse dispersion relation 
\be 
\omega _R^2\simeq {V_e^2\Omega _e^2k^2\over \Omega _e^2+V_e^2k^2}
\label{drt1}
\ee
where
\bdm 
\Omega _e={eB\over m_ec}
=1.76\cdot 10^7(B/{\rm Gauss}) \hbox{   Hz}
\edm
\be
=2.46\cdot 10^6\; (\epsilon _{B,-1}n_{b,8}\Theta _{-4})^{1/2}
\hbox{   Hz}
\label{electrongyro}
\ee
denotes the absolute value of the nonrelativistic
electron gyrofrequency and 
\be
V_e={\Omega _ec\over \sqrt{2}\omega _{p,e}}=
9.2\cdot 10^7(\epsilon_{B,-1}\Theta _{-4})^{1/2}
\; \hbox{ cm  s}^{-1}
\label{alfv1}
\ee
is the Alfven speed in the pair plasma.
In Eq. (\ref{drt1}) we adopt the convention 
that positive values of $\omega _R>0$ denote left-handed circularly polarized 
waves, and that negative values of $\omega _R<0 $ denote right-handed 
circularly polarized waves.

For small wavenumbers $|k|<<k_e$, where 
\be
k_e\equiv \Omega _e/V_e=2.7\cdot 10^{-2}n_{b,8}^{1/2}\; \hbox{ cm}^{-1}
\label{pairskin2}
\ee
is the inverse pair skin length, Eq. (\ref{drt1}) reduces to 
\be 
\omega _R^2=V_e^2k^2
\label{alfwavepair}
\ee
whereas in the opposite case $|k|>>k_e$ Eq. (\ref{drt1}) 
approaches the pair-cyclotron wave limit $\omega _R^2\simeq \Omega _e^2$.
For small wavenumbers ($|k|<<k_e$) we obtain for the pair Alfven waves
the damping rate
\be
\omega _{I,A}(k)\simeq 
-{\pi ^{1/2}\over 2}\; {\Omega _eV_e\over v_{th,e}}\; 
{k_e\over |k|}
\exp \Bigl[-({V_ek_e\over v_{th,e}k})^2\Bigr]
\label{dampingalfv1}
\ee
where $v_{th,e}=\sqrt{2\Theta }c=\sqrt{2k_BT_e/m_e}$ is the non-relativistic 
thermal pair velocity. 
\subsection{Time evolution of the electrostatic instability}
Neglecting spatial dependencies, 
the time-dependent behaviour of the intensities of the excited 
electrostatic waves
is given by
\be
{\partial I_e(k,t)\over \partial t}= 2\psi _e I_e(k,t),
\label{intense}
\ee 
where the growth rate $\psi _e$ is (Schlickeiser \cite{s02}, Ch. 11.2)
\bdm
\psi _e(k)\simeq 
{2\pi ^2\over \omega _R[{\partial \Re \Lambda _L\over \partial \omega _R}]}
\sum _{i=e,p}\omega _{p,i}^2 
\int_{-\infty}^\infty dp_{\parallel} \int_0^\infty 
dp_{\perp} {p_{\parallel}p_{\perp} \over \gamma }
\edm 
\bdm 
\delta ({kp_{\parallel}\over \gamma m_i}-\omega _R){\partial f_i \over \partial p_{\parallel}}
=\pi ^2\sum _{i=e,p}\omega _{p,i}^2
\int_{-\infty}^\infty dp_{\parallel} 
\edm 
\be 
\int_0^\infty dp_{\perp} {p_{\parallel}p_{\perp} \over \gamma }
\delta ({kp_{\parallel}\over \gamma m_i}-\omega _R)
{\partial f_i \over \partial p_{\parallel}}
\label{psie}
\ee
with
\be
\omega _{p,i}^2={4\pi n_ie_i^2\over m_i}
\label{omegapidef}
\ee 
Note that the summation $i=e,p$ refers to the incoming beam electrons and protons.
Eq. (\ref{intense}) holds provided that the initial growth rate $\psi _e(t=0)$ due to the 
beam particles is much larger than the Landau damping 
rate $\omega _I$ in the thermal pair plasma 
which is verified in Appendix B.

We do not know precisely the wave spectrum at time $t=0$. What we do know
is that one starts with a beam particle spectrum that is  a $\delta $-function in
momentum parallel to the ambient magnetic field and, once the instability stops,
the particle spectrum must be a plateau in $p_{\parallel}$. Thus the
instability rate at late times can be effectively calculated using the weak
turbulence limit. At early times such is not the case and more care has to be
taken (Pohl et al. \cite{pls02}). However, our interest here centers
on the late time evolution when the weak turbulence approximation is
particularly appropriate.

Integrating Eq. (\ref{intense}) over time gives
\bdm 
I_e(k,t)-I_e(k,t=0)=2\pi ^2
\sum _{i=e,p}\omega _{p,i}^2
\int_{-\infty}^\infty dp_{\parallel} \int_0^\infty 
dp_{\perp} 
\edm
\be
{p_{\parallel}p_{\perp} \over \gamma }
\delta ({kp_{\parallel}\over \gamma m_i}-\omega _R)
\int_0^tdt^{'}I(k,t^{'}){\partial f_i \over \partial p_{\parallel}}
\label{inte}
\ee
At the same time the phase space density of the beam particles evolves as
\be
{\partial f_i\over \partial t}=\pi e_i^2
{\partial \over \partial p_{\parallel }}\Bigl[ \int_{-\infty }^\infty 
dk\; I(k,t)\delta (\omega _R-{kp_{\parallel}\over \gamma m_i})
{\partial f_i\over \partial p_{\parallel }}\Bigr]
\label{fpe}
\ee
In terms of the phase space distribution function 
\be
f_i={1\over 2\pi p_{\perp}}\delta (p_{\perp})G_i(p_{\parallel },t)
\label{distzyl}
\ee
the two coupled Eqs. (\ref{inte}) and (\ref{fpe}) read 
\bdm 
I_e(k,t)-I_e(k,t=0)=\pi 
\sum _{i=e,p}\omega _{p,i}^2
\int_{-\infty}^\infty dp_{\parallel} 
{p_{\parallel}\over \sqrt{1+({p_{\parallel}\over m_ic})^2}}
\edm
\be
\delta ({kp_{\parallel}\over m_i\sqrt{1+({p_{\parallel}\over m_ic})^2}}
-\omega _R)
\int_0^tdt^{'}I(k,t^{'})
{\partial G_i (p_{\parallel }, t^{'})\over \partial p_{\parallel}}
\label{inte1}
\ee
and 
\bdm 
{\partial G_i\over \partial t}=\pi e_i^2
{\partial \over \partial p_{\parallel }}\bigl[ \int_{-\infty }^\infty \, dk
\; I(k,t)
\edm 
\be 
\delta (\omega _R-
{kp_{\parallel}\over m_i\sqrt{1+({p_{\parallel}\over m_ic})^2}})
{\partial G_i\over \partial p_{\parallel }}\bigr]
\label{fpe1}
\ee
Performing the $k$-integral in Eq. (\ref{fpe1}) with 
\bdm 
\delta (\omega _R-
{kp_{\parallel}\over m_i\sqrt{1+({p_{\parallel}\over m_ic})^2}})
=
\edm
\be 
{m_i\sqrt{1+({p_{\parallel}\over m_ic})^2}\over 
|p_{\parallel}|}\delta (k-
{m_i\omega _R\sqrt{1+({p_{\parallel}\over m_ic})^2}\over 
p_{\parallel}})
\label{delta1}
\ee
yields
\bdm 
{\partial G_i\over \partial t}=\pi e_i^2m_i
{\partial \over \partial p_{\parallel }}\bigl[
{\sqrt{1+({p_{\parallel}\over m_ic})^2}\over |p_{\parallel}|}
\edm 
\be 
I({m_i\omega _R\sqrt{1+({p_{\parallel}\over m_ic})^2}\over 
p_{\parallel}},t)
{\partial G_i\over \partial p_{\parallel }}\bigr]
\label{fpe2}
\ee
which can be integrated over $t$ and $p_{\parallel}$, resulting in
\bdm
\int_0^pdp_{\parallel}[G(p_{\parallel},t)-G(p_{\parallel},t=0)]=
\pi e_i^2m_i
{\sqrt{1+({p\over m_ic})^2}\over |p|}
\edm
\be
\int_0^tdt^{'}I({m_i\omega _R\sqrt{1+({p\over m_ic})^2}\over 
p},t^{'}){\partial G_i(p,t^{'})\over \partial p}
\label{fpe3}
\ee
We then obtain
\bdm
\int_0^tdt^{'}I({m_i\omega _R\sqrt{1+({p\over m_ic})^2}\over p},t^{'})
{\partial G_i(p,t^{'})\over \partial p}=
\edm
\be
{|p|\over 
\pi e_i^2m_i{\sqrt{1+({p\over m_ic})^2}}}
\int_0^pdp_{\parallel}[G(p_{\parallel},t)-G(p_{\parallel},t=0)]
\label{fpe4})
\ee
Likewise, we perform the $p_{\parallel}$-integration in the wave equation
(\ref{inte1}) using
\bdm 
\delta ({kp_{\parallel}\over m_i\sqrt{1+({p_{\parallel}\over m_ic})^2}}
-\omega _R)
=
\edm 
\be 
{m_i\over |k|}\bigl({N^2\over N^2-1}\bigr)^{3/2}
\delta (p_{\parallel}-{m_ic\; \sN \over \sqrt{N^2-1}})
\label{delta2}
\ee
where $N=ck/\omega _R$ is the index of refraction.
We obtain
\bdm
I_e(k,t)-I_e(k,t=0)=\pi c
\sum _{i=e,p}{\omega _{p,i}^2m_i^2\over |k|} 
{N^2\sN \over (N^2-1)^{3/2}}
\edm
\be
\int_0^tdt^{'}I(k,t^{'})
{\partial G_i (p_{\parallel }, t^{'})\over \partial p_{\parallel}}|
_{p_{\parallel}={m_ic\sN \over \sqrt{N^2-1}}}
\label{inte2}
\ee
We note that precisely the left-hand side of Eq. (\ref{fpe4}) taken at the 
values of $p={m_ic\sN \over \sqrt{N^2-1}}$ appears in the
wave equation (\ref{inte2}), so that after inserting Eq. (\ref{fpe4}) we derive
\bdm 
I_e(k,t)-I_e(k,t=0)=c^2\, {|N|\sN \over |k|(N^2-1)^{3/2}}
\sum _{i=e,p}{\omega _{p,i}^2m_i^2\over e_i^2} 
\edm
\be
\int_0^{m_ic\sN /\sqrt{N^2-1}}
dp_{\parallel}[G(p_{\parallel},t)-G(p_{\parallel},t=0)]
\label{quasi}
\ee
which relates electrostatic wave spectra and particle distribution functions
at time $t$ to the respective quantities at time $t=0$.
Before discussing the consequences of the quasilinear integral (\ref{quasi})
for electrostatic waves we derive the corresponding quasilinear integral
for transverse waves.
\subsection{Time evolution of the transverse instability}
Neglecting again spatial dependencies, 
the time-dependent behaviour of the intensities of the 
excited transverse waves is given by (Lee \& Ip \cite{li87}; 
Pohl \& Schlickeiser \cite{ps00})
\be
{\partial I_n\over \partial t}= 2\psi _n I_n,
\label{intens}
\ee 
where the growth rate $\psi _n$ is
\bdm
\psi _n(k)\simeq 
\Im {\pi\over \omega _R[{\partial \Re \Lambda _T\over \partial \omega _R}]}
\sum _{i=e,p}\omega _{p,i}^2
\int_{-\infty}^\infty dp_{\parallel} \int_0^\infty 
dp_{\perp} 
\edm 
\bdm 
{p^2_{\perp} \over \gamma 
({kp_{\parallel}\over \gamma m_i}+\Omega _i\gamma ^{-1}-\omega _R)}
[{\partial f_i \over \partial p_{\perp }}
+{k\over \gamma m_i\omega _R}(p_{\perp}
{\partial f_i \over \partial p_{\parallel}}
-p_{\parallel}{\partial f_i \over \partial p_{\perp}})]
\edm
\bdm
={\pi^2\over \omega _R[{\partial \Re \Lambda _T\over \partial \omega _R}]}
\sum _{i=e,p}\omega _{p,i}^2
\int_{-\infty}^\infty dp_{\parallel} \int_0^\infty 
dp_{\perp} {p^2_{\perp} \over \gamma }
\delta ({kp_{\parallel}\over \gamma m_i}
\edm
\be
+{\Omega _i\over \gamma }-\omega _R)
[{\partial f_i \over \partial p_{\perp }}
+{k\over \gamma m_i\omega _R}(p_{\perp}
{\partial f_i \over \partial p_{\parallel}}
-p_{\parallel}{\partial f_i \over \partial p_{\perp}})]
\label{transim}
\ee 
Eq. (\ref{intens}) holds provided that the initial growth rate $\psi _n(t=0)$ due to the 
beam particles is much larger than the cyclotron damping rate $\omega _{I,A}$ of the pair 
Alfven waves in the thermal 
background pair plasma which is verified in Appendix B.

Again, we operate in the weak turbulence limit so that the long-time behaviour
of the wave intensity can be used because, once again, we do not know the early
time evolution of the wave spectrum. 

Transforming from the momentum variables ($p_{\perp}, p_{\parallel}$)
to the new variables
\be 
y={p_{\parallel}\over m_ic},\;\; 
E=\gamma =\sqrt{1+{p^2_{\parallel}+p^2_{\perp}\over (m_ic)^2}}
\label{eytra}
\ee
we obtain
\bdm
\psi _n=
{\pi ^2\over |k|c\omega _R[{\partial \Re \Lambda _T\over \partial \omega _R}]}
\sum _{i=e,p}\omega _{p,i}^2(m_ic)^3\int_1^\infty dE
\edm
\be
\int_{-\sqrt{E^2-1}}^{\sqrt{E^2-1}}dy\, (E^2-1-y^2)
\delta (y-{E\over N}+x_i)({\partial f_i\over \partial E}+
N{\partial f_i\over \partial y})
\label{ima}
\ee
where
\be
x_i={\Omega _i\over kc}
\label{defxi}
\ee
For Alfven speeds $V_e$ that are much less than the speed of light $c$,
the index of refraction of $N=c/V_e>>1$ is large compared to unity,
so that the derivative $\partial f_i/\partial E$ is negligible compared
to $N(\partial f_i/\partial y)$, leading to
\bdm
\psi _n=
{\pi ^2\sk \over \omega ^2_R[{\partial \Re \Lambda _T\over \partial \omega _R}]}
\sum _{i=e,p}\omega _{p,i}^2(m_ic)^3
\edm
\be
\int_1^\infty dE\int_{-\sqrt{E^2-1}}^{\sqrt{E^2-1}}dy\, (E^2-1-y^2)
\delta (y-{E\over N}+x_i){\partial f_i\over \partial y}
\label{ima1}
\ee
Transforming from the variable $y$ to the cosine of the pitch angle $\mu =
y/\sqrt{E^2-1}$ gives
\bdm
\psi _n=
{\pi ^2\sk \over \omega ^2_R[{\partial \Re \Lambda _T\over \partial \omega _R}]}
\sum _{i=e,p}\omega _{p,i}^2 (m_ic)^3
\edm
\be
\int_1^\infty dE\sqrt{E^2-1}\int_{-1}^1d\mu \,
(1-\mu ^2)\delta (\mu -\mu _0){\partial f_i\over \partial \mu }
\label{ima2}
\ee
where
\be
\mu _0={EN^{-1}-x_i\over \sqrt{E^2-1}}
\label{mu0def}
\ee
The $\mu $-integration gives a nonvanishing value provided $|\mu _0|\le 1$
which is equivalent to the requirement $E\ge E_i$
where
\be 
E_i={\sqrt {1-N^{-2}+x_i^2}-x_iN^{-1}\over 1-N^{-2}}
\simeq \sqrt{1+x_i^2}
\label{eidef}
\ee
where the latter approximation holds in the limit $|N|>>1$.
Consequently, we derive
\bdm 
\psi _n=
{\pi ^2\sk \over \omega ^2_R[{\partial \Re \Lambda _T\over \partial \omega _R}]}
\sum _{i=e,p}\omega _{p,i}^2 (m_ic)^3
\edm
\be 
\int_{E_i}^\infty dE\sqrt{E^2-1}
(1-\mu _0^2){\partial f_i\over \partial \mu }|_{\mu =\mu _0}
\label{ima3}
\ee
In terms of the normalized phase space distribution function 
\be
f_i=\delta(E-\Gamma )F_i(\mu ,t)/[2\pi (m_ic)^3\Gamma (\Gamma^2 -1)^{1/2}]
\label{distn}
\ee
Eq. (\ref{ima3}) reads
\be
\psi _n=
{\pi \sk \over 2\omega ^2_R[{\partial \Re \Lambda _T\over \partial \omega _R}]}
\sum _{i=e,p}{\omega _{p,i}^2\over \Gamma }H[|k|-R_i^{-1}]
[1-\mu _i^2]
{\partial F_i\over \partial \mu }|_{\mu =\mu _i}
\label{grow}
\ee
where
\be
\mu _i={\Gamma N^{-1}-x_i\over \sqrt{\Gamma ^2-1}}=
{1\over kV}(\omega _R-{\Omega _i\over \Gamma })
\label{muidef}
\ee
$H$ denotes the step function, $R_i=V\Gamma /|\Omega _i|$ the gyroradius.

The assumption of very large $N=ck/\omega _R\to \infty $, made before, 
corresponds formally to $\omega _R\to 0$, so that we can approximate Eq. 
(\ref{muidef}) as
\be
\mu _i\simeq \mu _c=-{\Omega _i\over \Gamma kV}=-{\sq \over R_ik}
\label{muc}
\ee
According to Appendix A
\be 
\omega _R^2[{\partial \Re \Lambda _T\over \partial \omega _R}]=
{2c^2k^2\over \omega _R}+\; 
{4\omega ^2_{p,e}\omega _R^3\over (\omega _R^2-\Omega _e^2)^2}
\simeq 
{2c^2k^2\over \omega _R}
\label{waviden}
\ee
for frequencies much less than the electron gyrofrequency.
Inserting Eq. (\ref{waviden}) in Eq. (\ref{grow}) we obtain
\be
\psi _n=
{\pi \omega _R\over 4k|k|c^2}
\sum _{i=e,p}{\omega _{p,i}^2\over \Gamma }H[|k|-R_i^{-1}]
[1-\mu _c^2]
{\partial F_i\over \partial \mu }|_{\mu =\mu _c}
\label{grow1}
\ee
With the dispersion relation (\ref{alfwave}) we find for forward (+)
(i.e. $\omega _R=V_ek$) and backward (-) (i.e. $\omega _R=-V_ek$)
Alfven waves that the wave growth rates are
\be
\psi _{\pm }=\pm \psi 
\label{psipm}
\ee
with 
\be
\psi =
{\pi V_e\over 4|k|c^2\Gamma }
\sum _{i=e,i}\omega _{p,i}^2 H[|k|-R_i^{-1}]
[1-\mu _c^2]
{\partial F_i\over \partial \mu }|_{\mu =\mu _c}
\label{grow2}
\ee
and accordingly for the time evolution of the respective wave intensities
(see Eq. (\ref{intens}))
\be
{\partial I_+\over \partial t}= +2\psi  I_+,\;\;
{\partial I_-\over \partial t}= -2\psi  I_-
\label{intenspm}
\ee 
The two equations (\ref{intenspm}) yield the integrals 
\be
I_+(t)I_-(t)=I_+(t=0)I_-(t=0)
\label{int1}
\ee
and 
\be
[I_+(t)-I_-(t)]-[I_+(t=0)-I_-(t=0)]=Z(k)
\label{int2}
\ee
where
\bdm 
Z(k)=
{\pi V_e\over 2|k|c^2\Gamma }
\sum _{i=e,p}\omega ^2_{p,i} H[|k|-R_i^{-1}]
\edm 
\be 
\sum_{n=+,-}\int_0^tdt^{'}
[1-\mu _c^2]
{\partial F_i\over \partial \mu }|_{\mu =\mu _c}I_n(k,t^{'})
\label{zed}
\ee
The influence of these excited waves 
on the beam particles is described by the quasilinear Fokker-Planck
equation for the resonant wave-particle 
interaction. For Alfv\'en waves the index of refraction
is large compared to unity, so that the Lorentz force associated
with the magnetic field of the waves is much stronger than the Lorentz force
associated with the electric field. As a consequence, on the 
shortest time scale
these waves scatter the particles in pitch angle $\mu $ but conserve their
energy, i.e. the waves isotropise the beam particles. The Fokker-Planck 
equation for the evolution of the phase space density is then 
\be
{\partial F_i\over \partial t}=
{\partial \over \partial \mu }\bigl [
D_{\mu \mu }{\partial F_i\over \partial \mu }\bigr]\ , 
\label{fpt}
\ee
where the pitch angle Fokker-Planck coefficient is determined by 
the two wave intensities 
\bdm
D_{\mu \mu }= \sum _{n=+,-} {{\pi\,\Omega _i^2\,(1-\mu ^2)}\over
{2\,B^2\Gamma ^2}}
\edm 
\bdm 
\int_{-\infty}^{\infty} dk\, I_n(k,t)\delta 
(\omega _R-kV\mu -{\Omega _i\over \Gamma })\, 
\edm
\be
\simeq 
\sum _{n=+,-} {{\pi\,\Omega _i^2\,(1-\mu ^2)}\over
{2\,B^2\Gamma ^2}}
\int_{-\infty}^{\infty} dk\, I_n(k,t)\delta 
(kV\mu +{\Omega _i\over \Gamma })\,
\label{Dmumu}
\ee
where we again used the limit $\omega _R\to 0$.
Integrating Eq. (\ref{fpt}) over pitch angle and time, and using 
Eq. (\ref{muc}) we find
\bdm
\int _{-1}^{\mu }d\mu ^{'}[F_i(\mu ^{'},t)-F_i(\mu ^{'},t=0)]
={\pi \Omega _i^2\over 2\Gamma ^2B^2}\sum _{n=+,-}(1-\mu ^2)
\edm 
\bdm 
\int_{-\infty}^{\infty} dk\,
\int_0^tdt^{'}I_n(k,t^{'}){\partial F_i\over \partial \mu }
\delta (kV\mu +{\Omega _i\over \Gamma}) 
={\pi \Omega _i^2\over 2B^2\Gamma ^2V|\mu |}
\edm
\bdm 
\sum _{n=+,-}
\int_0^tdt^{'}(1-\mu ^2)I_n(-{\Omega _i\over \Gamma V\mu} ,t^{'})
{\partial F_i\over \partial \mu }
\edm
\be
=
{\pi \Omega _i^2\over 2B^2\Gamma ^2V|\mu |}\sum _{n=+,-}
\int_0^tdt^{'}(1-\mu ^2)I_n({\mu _ck\over \mu} ,t^{'})
{\partial F_i\over \partial \mu }
\label{int3}
\ee
Evaluating Eq. (\ref{int3}) at $\mu =\mu _c$ gives
\bdm
\sum_{n=+,-}\int_0^tdt^{'}I_{n}(k,t^{'})(1-\mu _c^2)
{\partial F_i\over \partial \mu }|_{\mu =\mu _c}=
\edm
\be
{2\Gamma ^2B^2V\over \pi \Omega _i^2}|\mu _c|
\int _{-1}^{\mu _c}d\mu ^{'}[F_i(\mu ^{'},t)-F_i(\mu ^{'},t=0)]
\label{int4}
\ee
which can be inserted into Eq. (\ref{zed}) to yield
\bdm
Z(k)=
{V_eB^2\over V|k|c^2\Gamma }\sum_{i=e,p}H[|k|-R_i^{-1}]\omega _{p,i}^2R_i^2
|\mu _c|
\edm 
\bdm 
\times \int _{-1}^{\mu _c}d\mu ^{'}[F_i(\mu ^{'},t)-F_i(\mu ^{'},t=0)]
\edm 
\bdm 
={V_eB^2\over Vk^2c^2\Gamma }\sum_{i=e,p}H[|k|-R_i^{-1}]\omega _{p,i}^2R_i
\edm 
\be 
\times \int _{-1}^{\mu _c}d\mu ^{'}[F_i(\mu ^{'},t)-F_i(\mu ^{'},t=0)]
\label{zed1}
\ee
The system of equations (\ref{int1}), (\ref{int2}), and (\ref{zed1}) 
has the general solution 
at time $t$ 
\bdm 
I_+(t)=\sqrt {Y+{1\over 4}(Z+I_+(0)-I_-(0))^2}
\edm 
\be 
+0.5\,\left(Z+I_+(0)-I_-(0)\right)
\label{iplus}
\ee
and
\bdm 
I_-(t)=\sqrt {Y+{1\over 4}(Z+I_+(0)-I_-(0))^2}
\edm 
\be 
-0.5\,\left(Z+I_+(0)-I_-(0)\right)
\label{iminus}
\ee
where
\be
Y\equiv I_+(0)I_-(0) 
\label{ydef}
\ee
If the initial turbulence is much weaker than the self-generated turbulence
$I(k,0)<<|Z(k)|$ and has a vanishing cross-helicity $I_+(k,0)=I_-(k,0)=I(k,0)$
we obtain for Eqs. (\ref{iplus},\ref{iminus}) approximately
\be
I_{\pm }(k,t)\simeq {1\over 2}[|Z|\pm Z]+\; {2I^2(k,0)\over |Z|}
\label{iplusminus}
\ee
which relates the transverse wave spectra and particle distribution functions
at time $t$ to the respective quantities at time $t=0$.
\section{Self-excited turbulence and initial beam relaxation}
We first discuss which of the two, electrostatic or
transverse, relaxation processes on the incoming interstellar electron-proton
beam is faster. Hence we
calculate the power spectra of the self-excited 
electrostatic and transverse turbulence from the derived quasilinear integrals
for the beam initial condition.

For both relaxation processes the  
initial boundary condition for the evolution is the same:
in the beginning
($t=0$) there is a mono-energetic beam distribution (\ref{dist_f1}), i.e.
in terms of the two normalised distributions (\ref{distzyl}) and (\ref{distn})
\bdm 
G_{p,e}(p_{\parallel},t=0)=\delta (p_{\parallel}+P),
\edm 
\be 
F_{p,e}(\mu ,t=0)=\delta (\mu +1) 
\label{inif}
\ee
The final state of the evolution of the electrostatic instability is reached
at time $t_e$ when both growth rate and temporal
derivative of the particle distribution function are zero, i.e. when
$\partial G_i/\partial p_{\parallel }=0$ which is referred to as the
"plateau distribution"
\be
G_i(p_{\parallel },t_e)={1\over P}H[p_{\parallel }+P]H[-p_{\parallel }]
\label{finalg}
\ee
The final state of the evolution of the transverse instability 
is reached at time $t_t$ when both growth rate and temporal
derivative of the distribution are zero, i.e. when 
$\partial F_i/\partial \mu =0$. Consequently here
\be
F_{i}(\mu , t=t_t)={1\over 2} 
\label{finalf}
\ee
At this time the magnetohydrodynamic waves have completely isotropised the 
beam distribution.
\subsection{Self-excited electrostatic turbulence}
Inserting Eqs. (\ref{inif}) and (\ref{finalg}) in Eq. (\ref{quasi}) allows us 
to integrate over $p_{\parallel }$ to obtain
\bdm
\sN \int_0^{m_ic\sN /\sqrt{N^2-1}}
dp_{\parallel}[G(p_{\parallel},t_e)-G(p_{\parallel},t=0)]
\edm
\be 
={H[-N]\over \sqrt{N^2-1}\sqrt{\Gamma ^2-1}}
H[\Gamma -{|N|\over \sqrt{N^2-1}}]
\ee
so that the fully developed electrostatic turbulence is
\bdm
I_e(k,t_e)-I_e(k,t=0)={c^2\over \sqrt{\Gamma ^2-1}}
{|N|H[-N]\over |k|(N^2-1)^2}
\edm 
\bdm 
H[\Gamma -{|N|\over \sqrt{N^2-1}}]
\sum _{i=e,p}{\omega _{p,i}^2m^2_i\over e_i^2}
\edm
\be
={4\pi n_im_pc^2\over \sqrt{\Gamma ^2-1}}(1+{m_e\over m_p}) 
{|N|H[-N]\over |k|(N^2-1)^2}H[\Gamma -{|N|\over \sqrt{N^2-1}}]
\label{quasi1}
\ee
The total enhancement in electric field fluctuation power due to the
plateauing of the proton and electron distribution function is obtained 
by integrating Eq. (\ref{quasi1})
\bdm 
(\delta E)^2=\int_{-\infty}^0 dk[I_e(k,t_e)-I_e(k,t=0)]
\edm
\bdm 
={4\pi n_im_pc^2\over \sqrt{\Gamma ^2-1}}(1+{m_e\over m_p})
\int_{(1-\Gamma ^{-2})^{-1/2}}^\infty {dN \over (N^2-1)^2}
\edm
\bdm
={2\pi n_im_pc^2\over \sqrt{\Gamma ^2-1}}(1+{m_e\over m_p})
\bigl[\Gamma \sqrt{\Gamma ^2-1}-\; 
{1\over 2}\ln {\Gamma +\sqrt{\Gamma ^2-1}\over \Gamma -\sqrt{\Gamma ^2-1}}
\bigr]
\edm
\be
\simeq 2\pi \Gamma n_im_pc^2(1+{m_e\over m_p})
\label{totalefield}
\ee
so that the change in the electric field fluctuation energy density is
\be
\Delta U_{(\delta E)}={(\delta E)^2\over 8\pi }={1\over 4}\Gamma n_im_pc^2
(1+{m_e\over m_p})
\label{efenerg}
\ee
Electrostatic waves possess equipartition between electric and velocity
fluctuations. so that the total change in fluctuation energy density due to
plateauing is
\be
\Delta U=2\Delta U_{(\delta E)}=
{1\over 2}\Gamma n_im_pc^2(1+{m_e\over m_p})
\label{efenerg1}
\ee
which is half of the initial energy density of the beam particles.
\subsection{Electrostatic beam relaxation}
Using the fully developed turbulence spectrum (\ref{quasi1}) we obtain for
the Fokker-Planck equation (\ref{fpe2}) in the variable 
$x=|p_{\parallel }|/m_ic$ 
\be
{\partial G_i\over \partial t}=
{\partial \over \partial x}\bigl[{x^2\over \tau _i(x)}
{\partial G_i\over \partial x}\bigr]
\label{fpefullx}
\ee
with the time scale 
\be
\tau _i(x)
={\sqrt{2(\Gamma ^2-1)}\over \pi (1+{m_e\over m_p})}
{m_i^2\over m_em_p}{n_b\over n_i}\omega ^{-1}_{p,e}
{1\over x\sqrt{1+x^2}}
\label{taui}
\ee
for values of  $0\le x\le \sqrt {\Gamma^2-1}$.

Evaluating this time scale at the momentum $x=P/m_ic$
yields for the beam electrons
\be
t_{e,e}={\sqrt{2}\over \pi \omega _{p,e}}{m_e\over m_p(1+{m_e\over m_p})}
{n_b\over n_i\Gamma }
\label{taue}
\ee
and for the beam protons
\be
t_{e,p}=({m_p\over m_e})^2t_{e,e}
\label{taup}
\ee

\subsection{Self-excited transverse turbulence}
Inserting Eqs. (\ref{inif}) and (\ref{finalf}) in Eq. (\ref{zed1}) allows us
to integrate over $\mu $ to obtain with
\be
\int _{-1}^{\mu _c}d\mu ^{'}[F_i(\mu ^{'},t_t)-F_i(\mu ^{'},t=0)]=
-{1\over 2}[1-\mu _c]
\label{deltafint}
\ee
so that
\bdm
Z(k)=-
{V_eB^2\over 2Vk^2c^2\Gamma }\sum_{i=e,p}
H[|k|-R_i^{-1}]\omega _{p,i}^2R_i
[1+{\sq \over R_ik}]
\edm
\bdm 
=-{V_eB^2R_p\omega _{p,e}^2\over 2Vk^2c^2}{n_i\over n_b\Gamma }{m_e\over m_p}
\Bigl(
H[|k|-R_p^{-1}](1+{1\over R_pk}]+\; 
\edm 
\be 
H[|k|-{m_p\over m_e}R_p^{-1}](1-{(m_p/m_e)\over R_pk}]\Bigr)
\label{azed1}
\ee
which is negative 
for all wavenumber values. According to Eq. (\ref{iplusminus})
this implies that predominantly backward moving left-handed and right-handed 
polarized Alfven waves are generated. If the initial turbulence is negligibly
small the 
respective intensities of the fully developed turbulence are 
\be
I_{-}(k,t_t)=|Z(|k|)|,\; \;\,\;
I_+(k,t_t)=0
\label{resulint}
\ee
The total enhancement in magnetic field fluctuation power due to the
isotropisation of the proton and electron distribution function is obtained 
by integrating Eq. (\ref{int2}) using Eq. (\ref{resulint})
\bdm
(\delta B)_-^2=\int_{-\infty}^\infty dk\sum _{n=\pm }
[I_n(k,t_t)-I_n(k,t=0)]=
\edm 
\bdm 
-\int_{-\infty }^\infty dk Z(k)=
{m_e\over m_p}{V_e\over V}{n_i\over n_b\Gamma }
{B^2R_p^2\omega _{p,e}^2\over c^2}[1+{m_e\over m_p}]
\edm 
\be
=4\pi n_i\Gamma m_pV_eV[1+{m_e\over m_p}]
\label{totalb}
\ee
so that the change in the magnetic field fluctuation energy density is
\be
\Delta U_{\delta B}={(\delta B)_-^2\over 8\pi }=
{1\over 2}n_im_pV_eV\Gamma [1+{m_e\over m_p}]
\label{totalub}
\ee
Alfven waves possess equipartition of wave energy densities between magnetic and
plasma velocity fluctuations, so that the total change in fluctuation energy
density to pitch angle isotropisation is 
\be
\Delta U=\Delta U_{\delta B}+\Delta U_{\delta v}=2\Delta U_{\delta B}=
n_im_pV_eV\Gamma [1+{m_e\over m_p}]
\label{totalu}
\ee
Following exactly the same
calculation as in Pohl \& Schlickeiser (\cite{ps00}) one can 
demonstrate that the increase (\ref{totalu}) in the
energy density is balanced by a corresponding decrease in the energy density of
the beam protons and electrons during their isotropisation. The transverse
plasma turbulence is generated at the expense of the beam particles which relax
to a state of lower energy density. The change in the energy density of the beam
particles is given by $-\Delta U$ which is the fraction
$V_e/c$ of the initial beam energy density.

\subsection{Transverse beam relaxation}

We can estimate the isotropisation length of the beam particles by using
the fully-developed turbulence spectrum 
of backward moving transverse waves (\ref{azed1})-(\ref{resulint})
\be
I_{-}(k,t_t)\simeq i_-|k|^{-2}H[|k|-R_p^{-1}]
\label{iminful1}
\ee
with
\be
i_-={V_eB^2R_p\omega _{p,e}^2\over 2Vk^2c^2}
{n_i\over n_b\Gamma }{m_e\over m_p}
\label{imin1}
\ee
to calculate 
the pitch angle Fokker-Planck coefficient (\ref{Dmumu})
\bdm
D_{\mu \mu }= {{\pi\Omega _i^2(1-\mu ^2)}\over
{2B^2\Gamma ^2}}
\int_{-\infty}^{\infty} dk\, I_-(k,t_t)\delta 
(\omega _R-kV\mu -{\Omega _i\over \Gamma })=
\edm
\be
d_i(1-\mu ^2)\int_{-\infty}^{\infty} dk\, |k|^{-2}H[|k|-R_p^{-1}]
\delta (\omega _R-kV\mu -{\Omega _i\over \Gamma })
\label{dmut1}
\ee
where
\be
d_i={\pi \over 4}{\Omega _i^2\over \Gamma ^2}
{V_e\omega _{p,e}^2\over \Omega _pc^2}{n_i\over n_b}{m_e\over m_p}
\label{d0def}
\ee
The cyclotron resonance condition of beam particles of species $i=e,p$
appearing in the argument of the $\delta $-function in Eq. (\ref{dmut1})
\be
\omega_R - k V \mu = \Omega_i/\Gamma,
\label{cr1}
\ee
when combined with the pair-plasma dispersion relation (with $V_e^2\ll
c^2$) for backward moving waves,
\be
\omega _R= -V_e k \sqrt {\frac{\Omega_e^2}{\Omega_e^2+k^2 V_e^2}},
\label{drback}
\ee 
gives the particle pitch angle cosine as a function of 
the resonant wavenumber in
the form
\be 
\mu _i(k) = -\epsilon \Bigl[
\sqrt {\Omega_e^2\over \Omega_e^2 + k^2 V_e^2} + \; 
{\Omega_i\over kV_e\Gamma }\Bigr]
\label{mures}
\ee
where we introduced 
\be
\epsilon \equiv V_e/V
\label{defeps}
\ee
In our convention, positive [negative] $k$ corresponds to right [left]
handed polarization. 

It is convenient to introduce the dimensionless wavenumber $x=k/k_e$ and
\be
\gamma _i\equiv {\Gamma _i|\Omega _e|\over \Omega _i}
\label{defgami}
\ee
so that
\be
D_{\mu \mu }= 
{d_i\over V_ek_e^2}(1-\mu ^2)\int_{-\infty}^{\infty} dx\, 
x^{-2}H[|x|-\epsilon \gamma _p^{-1}]
\delta (xg_i(x))
\label{dmut2}
\ee
with
\be
g_i(\mu ,x)=
{\mu \over \epsilon }+(1+x^2)^{-1/2}+(\gamma _ix)^{-1}
\label{gi1}
\ee
In terms of the dimensionless notation the resonance
condition (\ref{cr1}) then is equivalent to $g_i(\mu ,x)=0$ yielding
\be
{\mu (x)\over \epsilon }=-\bigl[(1+x^2)^{-1/2}+\, (\gamma _ix)^{-1}\bigr]
\label{mures1}
\ee
As a consequence of pitch-angle scattering 
the beam particles adjust to the isotropic distribution
on a length scale given by the scattering length 
\be
\lambda ={3V\over 8}\int_{-1}^1d\mu {(1-\mu ^2)^2\over D_{\mu \mu }(\mu )}
\label{scat1}
\ee
Eq. (\ref{scat1}) is valid for scattering lengths $\lambda $ larger than the
gyroradius of the particles, $R_p=3\cdot 10^9\Gamma _{3}/B$ cm and
$R_e=1.7\cdot 10^6\Gamma _{3}/B$ cm for protons and electrons, respectively. 
If the thickness $d$ of the outflow region is larger than 
the scattering length, indeed an isotropic distribution
of the inflowing interstellar protons and electrons with Lorentzfactor
$<\Gamma >=\Gamma (1-{V_eV\over c^2})\simeq \Gamma $
in the blast wave frame 
is efficiently generated. 
 
From the scattering length we derive the isotropisation time scale in the 
outflow plasma
\be
t_t=\lambda /c
\label{transversetime}
\ee 
\subsubsection{Beam proton cyclotron resonance in a pair plasma}
For beam protons
\be
\gamma _p=m_p\Gamma /m_e>>1
\label{gamp}
\ee
is large and positive. According to Eq. (\ref{mures1}) for protons, 
$\mu _p(x)$ is
a monotonically increasing function for $x>0$ going from $-\infty$ to
$0$ as $x$ goes from $0$ to $\infty$. For $x<0$,
$\mu _p \to\infty$ as $x\to 0-$, but as $-x$ increases,
$\mu _p$ has a minimum at
\be
x_E=-\gamma _p^{-1/3}[1-\gamma _p^{-2/3}]^{-1/2}\simeq 
-\gamma _p^{-1/3}
\label{xpextr}
\ee
and $\mu _p=0$ at
\be
x_0=-(\gamma _p^2-1)^{-1/2}\simeq -\gamma _p^{-1}
\label{xp0}
\ee
The minimum value is
\be
\mu _E=-\epsilon [1-\gamma _p^{-2/3}]^{3/2}\simeq 
-\epsilon [1-{3\over 2}\gamma _p^{-2/3}]
\simeq -\epsilon 
\label{muminp}
\ee
Because this value is negative, there is no resonance gap for protons
in a pair plasma even in a case of degenerate cross helicity. 
This is in contrast with the situation in a cold
electron--proton plasma, where one has to rely on electron--cyclotron
waves scattering ions if waves propagating in one direction only are
considered (see Vainio \cite{Vain00} for a complete analysis of the
resonance conditions in that case).

The $x$-value of the right-handed wave in resonance with the extreme
value $\mu _E$ achievable by the left-handed waves is approximately
\be
x\simeq  2\gamma _p^{-1/3}
\label{xrhp}
\ee
which is always much less than unity, indicating that we are well
in the non-dispersive range of the dispersion relation of the right-handed 
waves.
Thus, as far as phase speeds are considered, we can regard all the
resonant waves as Alfv\'enic. This is because dispersion modifies the
resonance conditions enabling relativistic particles to interact with
the Alfv'en waves over the whole momentum space. 
\begin{figure}
%\vskip-0.1in
\resizebox{\hsize}{!}{\includegraphics{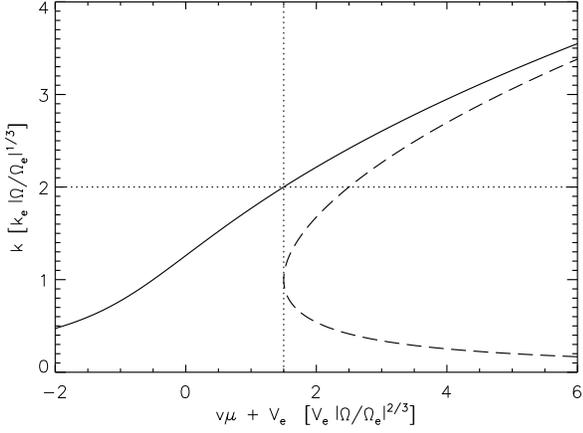}}
\caption{Resonant wavenumber as a function of parallel momentum near
the Alfv\'enic resonance gap. Here, $\Omega=\Omega_i/\Gamma$, $v=V$, 
the solid [dashed] curve gives $k$ for the right-handed [left-handed]
waves for ions and vice-versa for electrons.}
\label{fig1}
\end{figure}
We have plotted the resonant
wavenumber as a function of $V\mu$ close to $V\mu=-V_e$ in
Fig. 1. The figure is plotted for protons with
$\Gamma=100$. 
\subsubsection{Beam electron cyclotron resonance in a pair plasma}
For relativistic beam electrons ($\gamma _e=-\Gamma _e=-\Gamma < 0$ is a 
large negative
number), the analysis is very similar. The only
difference is that $x$ has to change sign relative to the proton case. Now one 
has a monotonic resonance function, $\mu _e(x)$, with
the left-handed waves, and the turning point in $\mu _e(x)$ for
right-handed waves.  
$\mu _e$ has a minimum at
\be
x_E=\Gamma _e^{-1/3}[1-\Gamma _e^{-2/3}]^{-1/2}\simeq 
\Gamma _e^{-1/3}
\label{xeextr}
\ee
and $\mu _e=0$ at
\be
x_0=(\Gamma _e^2-1)^{-1/2}\simeq \Gamma _e^{-1}
\label{xe0}
\ee
The minimum value is
\be
\mu _E=-\epsilon [1-\Gamma _e^{-2/3}]^{3/2}\simeq 
-\epsilon [1-{3\over 2}\Gamma _e^{-2/3}]\simeq -\epsilon 
\label{mumine}
\ee

Having established that no resonance gaps exist in the quasilinear
interaction of the particles and the backward moving waves of
relatively low wavenumbers, we will assume that other wave modes do
not affect the dynamics of the system. In this case we can evaluate
the wave spectra also with an alternative bi-spherical shell method
which is described in Appendix A.
\subsubsection{Scattering length of beam protons}
The function $g_i(\mu ,x)$ appearing in the 
Fokker-Planck coefficient (\ref{dmut2}) for beam protons is
\be
g_p(\mu ,x)=
{\mu \over \epsilon }+(1+x^2)^{-1/2}+(\gamma _px)^{-1}
\label{gp1}
\ee
According to the discussion of the proton cyclotron resonance condition
we approximate the function $g_p(\mu ,x)$ as
\be
g_p(\mu ,x)\simeq 
\cases{
{\mu \over \epsilon}+{1\over x\gamma _p}& for $-1\le \mu \le -\epsilon $\cr
{\mu \over \epsilon}-{\gamma _px+1\over \gamma _p^{2/3}-1}& 
for $-\epsilon \le \mu \le 0$\cr 
{\mu \over \epsilon}+1+{1\over \gamma _px}& for $0\le \mu \le 1$\cr}
\label{gpapp}
\ee
allowing a straightforward approximation of the proton Fokker-Planck
coefficient
\bdm 
D_{\mu \mu}^p(\mu )\simeq {d_p\gamma _p^2(1-\mu ^2)\over V_ek_e^2}
\edm 
\be 
\cases{
{|\mu |\over \epsilon }& for $-1\le \mu \le -\epsilon $\cr
(\gamma _p^{2/3}-1)|(\gamma _p^{2/3}-1){\mu \over \epsilon }-1|^{-3}& 
for $-\epsilon \le \mu \le 0 $\cr 
|{\mu \over \epsilon }+1|& for $0\le \mu \le 1$\cr }
\label{dmup1}
\ee
Inserting Eq. (\ref{dmup1}) into Eq. (\ref{scat1}) gives the proton scattering
length 
\bdm
\lambda _p={3VV_ek_e^2\over 8d_p\gamma _p^2}
\Bigl[\epsilon \int_{-1}^{-\epsilon }d\mu {(1-\mu ^2)\over |\mu |}
\edm 
\bdm 
+{1\over \gamma _p^{2/3}-1}\int_{-\epsilon }^0d\mu (1-\mu ^2)
|(\gamma _p^{2/3}-1){\mu \over \epsilon }-1|^3+
\edm
\bdm
\int_0^1d\mu {(1-\mu ^2)\over |1+{\mu \over \epsilon }|}\Bigr]
\simeq {3V_e^2k_e^2\over 8d_p\gamma _p^2}\Bigl[
2\ln \epsilon ^{-1}+ {(\gamma _p^{2/3}-1)^2\over 4}-1\Bigr]
\edm
\be
\simeq {3\over 32}{V_e^2k_e^2\over d_p\gamma _p^{2/3}}
={3\sqrt{2}\over 8\pi }{c\over \omega _{p,e}}{n_b\over n_i}
({m_p\Gamma \over m_e})^{4/3}
\label{scatp1}
\ee
The corresponding proton isotropisation time scale is
\be
t_{t,p}={\lambda _p\over c}=
{3\sqrt{2}\over 8\pi }{1\over \omega _{p,e}}{n_b\over n_i}
({m_p\Gamma \over m_e})^{4/3}
\label{tTp}
\ee
Using the proton resonance approximation (\ref{gpapp}) we obtain the smallest
resonant wavenumbers at $\mu =\pm 1$  from the condition
$g_p(\mu =\pm 1,x)=0$ as
\be 
x_{p,min}\simeq \mp {\epsilon \over \gamma _p}
\label{xpmin}
\ee
which corresponds to
\be 
|k_{p,min}|=k_e|x_{p,min}|={m_e\over m_p\Gamma }{\Omega _e\over v}=
{1\over R_p}
\label{kpmin}
\ee
This indicates that the beam protons resonate with Alfven waves with 
wavenumbers larger than the inverse of the proton gyroradius.
\subsubsection{Scattering length of beam electrons}
The function $g_i(\mu ,x)$ appearing in the 
Fokker-Planck coefficient (\ref{dmut2}) for beam electrons is
\be
g_e(\mu ,x)=
{\mu \over \epsilon }+(1+x^2)^{-1/2}-(\Gamma _ex)^{-1}
\label{ge1}
\ee
which we approximate as
\be
g_e(\mu ,x)\simeq 
\cases{
{\mu \over \epsilon}-{1\over x\Gamma _e}& for $-1\le \mu \le -\epsilon $\cr
{\mu \over \epsilon}+{\Gamma _ex-1\over \Gamma _e^{2/3}-1}& 
for $-\epsilon \le \mu \le 0$\cr 
{\mu \over \epsilon}+1-{1\over \Gamma _ex}& for $0\le \mu \le 1$\cr}
\label{geapp}
\ee
The electron Fokker-Planck coefficient then is approximated as
\bdm 
D_{\mu \mu}^e(\mu )\simeq {d_e\Gamma _e^2(1-\mu ^2)\over V_ek_e^2}
\edm 
\be 
\cases{
{|\mu |\over \epsilon }& for $-1\le \mu \le -\epsilon $\cr
(\Gamma _e^{2/3}-1)|(\Gamma _e^{2/3}-1){\mu \over \epsilon }-1|^{-3}& 
for $-\epsilon \le \mu \le 0 $\cr 
|{\mu \over \epsilon }+1|& for $0\le \mu \le 1$\cr }
\label{dmue1}
\ee
Inserting Eq. (\ref{dmue1}) into Eq. (\ref{scat1}) gives the electron scattering
length 
\be
\lambda _e\simeq {3\over 32}{V_e^2k_e^2\over d_e\Gamma _e^{2/3}}
={3\sqrt{2}\over 8\pi }{c\over \omega _{p,e}}{n_b\over n_i}
\Gamma ^{4/3}
\label{scate1}
\ee
and the corresponding electron isotropisation time scale 
\be
t_{t,e}={\lambda _e\over c}=
{3\sqrt{2}\over 8\pi }{1\over \omega _{p,e}}{n_b\over n_i}
\Gamma ^{4/3}
\label{tTe}
\ee
Using the electron resonance approximation (\ref{geapp}) we obtain the smallest
resonant wavenumbers at $\mu =\pm 1$  from the condition
$g_e(\mu =\pm 1,x)=0$ as
\be 
x_{e,min}\simeq \pm {\epsilon \over \Gamma _e}
\label{xemin}
\ee
which corresponds to
\be 
|k_{e,min}|=k_e|x_{e,min}|={\Omega _e\over v\Gamma }=
{1\over R_e}
\label{kemin}
\ee
The beam electrons resonate with Alfven waves with wavenumbers
larger than the inverse of the electron gyroradius. Therefore, isotropisation 
of the beam electrons is possible with their own self-generated turbulence, 
even in cases 
where the beam protons would not generate transverse turbulence.
\subsection{Electrostatic or transverse beam realaxation: which is faster ?}
By comparing the respective time scales, Eqs. (\ref{taue}), (\ref{taui}), 
(\ref{tTe}) and (\ref{tTp}), we can answer the important question
which beam relaxation process in pair plasmas is faster; plateauing in 
parallel momentum
by the electrostatic instability or pitch-angle isotropisation by the 
transverse instability.

For the beam electrons we obtain the ratio of relaxation times
\be
{t_{e,e}\over t_{t,e}}={8\over 3}{m_e\over m_p}\Gamma ^{-7/3}
<<1
\label{tsre}
\ee
while for the beam protons
\be
{t_{e,p}\over t_{t,p}}={8\over 3}({m_e\over m_p})^{1/3}\Gamma ^{-7/3}
<<1
\label{tsrp}
\ee
In both cases the electrostatic relaxation time is much smaller than 
the transverse relaxation time, so that the incoming interstellar protons
and electrons first relax to the plateau distribution (\ref{finalg}) in 
parallel momentum on the time sales $t_{e,e}$ and $t_{e,p}$. As a consequence,
the initial energy density of the beam particles reduces to half of its value; 
the other half has been transformed to electric field fluctuations of the
excited electrostatic turbulence (see Eq. (\ref{efenerg1}).

We can now study the secular evolution after time $t_{e,e}$
and $t_{e,p}$, respectively, of the plateau distribution 
(\ref{finalg}) which is still unstable with respect to the excitation of
transverse waves.

\section{Secular transverse beam relaxation}
We again use the derived quasilinear integrals for transverse beam evolution
during the secular phase. Here the boundary conditions are modified:
at the beginning of the secular phase 
($t=t_e$) the interstellar electrons and protons 
obey the plateau distribution 
\be
G_i(p_{\parallel },t_e)={1\over P}H[p_{\parallel }+P]H[-p_{\parallel }]
\label{secini1}
\ee
corresponding to
\be 
f_i(p,\mu ,t_e)={1\over 2\pi Pp^2}\delta (\mu +1)H[P-p]
\label{secini2}
\ee
Again the final state of the evolution of the transverse instability 
is reached at time $t_s$ when both growth rate and temporal
derivative of the distribution are zero, i.e. when 
$\partial F_i/\partial \mu =0$, corresponding
to
\be
f_{i}(p,\mu , t=t_s)={H[P-p]\over 4\pi Pp^2}
\label{secf}
\ee
At this time the magnetohydrodynamic waves have completely isotropised the 
plateau distribution.

Starting from Eqs. (\ref{intens}) and (\ref{ima3}) we can repeat the earlier
analysis mutatis mutandi to derive instead of Eq. (\ref{azed1}) that in this
case
\bdm
Z(k)={2\pi V_eB^2\over |k|c^2}\sum_i{\omega _{p,i}^2\over \Omega _i^2}
\int_{m_ic|x_i|}^\infty dpp^2v\sqrt{1+{p^2\over m_i^2c^2}}|\mu _c|
\edm 
\bdm 
\int_{-1}^{\mu _c}d\mu [f_i(p,\mu ,t_s)-f_i(p,\mu ,t_e)]
={2\pi V_eB^2\over k^2c^2}\sum_i{\omega _{p,i}^2\over |\Omega _i|}
\edm 
\be 
\int_{m_i|\Omega _i/k|}^\infty dpp^2
\int_{-1}^{\mu _c}d\mu [f_i(p,\mu ,t_s)-f_i(p,\mu ,t_e)]
\label{tzed1}
\ee
Using Eqs. (\ref{secini2})-(\ref{secf}) we obtain
\be
\int_{-1}^{\mu _c}d\mu [f_i(p,\mu ,t_s)-f_i(p,\mu ,t_e)]
=-{1-\mu _c\over 4\pi p^2P}H[P-p]
\label{delfs}
\ee
which after momentum integration yields for Eq. (\ref{tzed1})
\bdm
Z(k)=-{V_eB^2\over 2k^2c^2}\sum_i{\omega _{p,i}^2\over |\Omega _i|}
[1-{1\over |k|R_i}+{\sq \over kR_i}\ln (|k|R_i)]
\edm 
\bdm 
H[|k|-R_i^{-1}]
=-{V_eB^2\over 2k^2c^2V\Gamma }\sum_i\omega _{p,i}^2R_i
\edm 
\be 
[1-{1\over |k|R_i}+{\sq \over kR_i}\ln (|k|R_i)]H[|k|-R_i^{-1}]
\label{tzed2}
\ee
which also results from applying the bi-spherical shell method of Appendix C
to the flat distribution (\ref{secini2}). For large wavenumbers $|k|R_p>>1$ Eq.
(\ref{tzed2}) is well approximated by 
\be
Z(k)\simeq -{V_eB^2\over 2k^2c^2V\Gamma }\sum_i\omega _{p,i}^2R_i
\label{tzed3}
\ee
which is identical to the previous result
(\ref{azed1}). Again $Z(k)$ is negative 
for all wavenumber values, implying according to Eq. (\ref{iplusminus})
that predominantly backward moving left-handed and right-handed 
polarized Alfven waves are generated. If the transverse turbulence 
at time $t_e$ is negligibly
small the 
respective intensities of the fully developed turbulence 
at the end of the secular evolution phase are
\be
I_{-}(k,t_s)=|Z(|k|)|,\; \;\,\;
I_+(k,t_s)=0
\label{resulintsec}
\ee
The generated wavenumber spectrum and the total change in transverse wave
energy density is independent whether one starts
from a monoenergetic beam (as we did in the initial phase of the evolution)
or from a plateau distribution resulting from the faster electrostatic
relaxation (as we did in the secular phase). This implies that we get exactly
the same time scales for the pitch-angle isotropisation, 
(\ref{tTp}) and (\ref{tTe}), of the plateau distribution function, so that at
the end of the secular phase, $t_s=t_{t,e}$ for electrons and $t_s=t_{t,p}$
for protons, 
the swept-up interstellar protons and electrons have a flat and isotropic 
differential momentum spectrum
\be
N_i(p,t_s)=4\pi p^2f_{i}(p,\mu , t=t_s)={H[P-p]\over  P}
\label{dnfins}
\ee
in the rest frame of the plasma blob.
\section{Application to AGN jet outflows}
In order for the electrostatic and the transverse instability to plateau 
and isotropise the incoming interstellar electron and proton beam distributions,
the respective co-moving time scales $t_e$ and $t_t$ have to be smaller than the
crossing time of the beam particles which is given by the co-moving 
light crossing time of the blob 
\be
t_l=d/c=3\cdot 10^4d^*_{13}\Gamma _2 \hbox{   s}
\label{lightcross}
\ee
We consider these conditions separately for beam plateauing and beam
isotropisation.
\subsection{Beam plateauing}
According to Eqs. (\ref{taue}) and (\ref{lightcross}) for beam electrons 
the requirement
$t_{e,e}<t_{l}$ leads to
\bdm
{\sqrt{2}\over \pi \omega _{p,e}}{m_e\over m_p}
{n^*_b\over n_i^*\Gamma ^3}<3\cdot 10^4d^*_{13}\Gamma _2
\edm
which is identical to
\be
{(n^*_{b,10})^{1/2}\over n_i^*d^*_{13}\Gamma _2^{7/2}}<6.9\cdot 10^{12}
\label{coneplat}
\ee
which is well fulfilled for standard AGN jet outflow and environment parameters.

Likewise with Eq. (\ref{taup}) for beam protons the requirement 
$t_{e,p}<t_l$ becomes
\be
{(n^*_{b,10})^{1/2}\over n_i^*d^*_{13}\Gamma _2^{7/2}}<2.0\cdot 10^6
\label{conpplat}
\ee
which is also fulfilled for standard jet outflow and environment parameters.

We therefore conclude that for standard AGN jet outflow and environment parameters 
the initial beam distributions of interstellar protons and electrons relax 
to plataeu distributions in parallel momentum, transferring thereby one-half 
of the initial energy density of the beam particles to electric field
fluctuations of the generated electrostatic turbulence.
\subsection{Beam isotropisation and generation of radiation}
According to Eqs. (\ref{tTe}) and (\ref{lightcross}) for plateaued electrons 
the requirement
$t_{t,e}<t_{l}$ leads to
\be
{(n^*_{b,10})^{1/2}\over n_i^*d^*_{13}\Gamma _2^{7/6}}< 10^8
\label{coneiso}
\ee
which is fulfilled for standard jet outflow and environment 
parameters.

For plateaued protons Eqs. (\ref{tTp}) and (\ref{lightcross}) imply 
\be
{(n^*_{b,10})^{1/2}\over n_i^*d^*_{13}\Gamma _2^{7/6}}<4.4\cdot 10^3
\label{conpiso}
\ee
which is also fulfilled for the adopted standard jet outflow and 
environment parameters.

If we take the adopted AGN jet outflow ($n_{b,10}^*=1$, $\Gamma _2=1$, 
$d^*_{13}=1$) and 
environment ($n_i^*$) parameters as face-value parameters, we conclude that 
the interstellar electrons and protons will isotropise by their self-generated
transverse turbulence and thus be picked-up by the outflow pair plasma.
In Appendix D we demonstrate that both transverse wave isotropisation time scales 
(\ref{tTe}) and (\ref{tTp}) are much shorter than the isotropisation of beam 
electrons and protons by elastic Coulomb interactions with pair jet plasma.

In the pair outflow frame the external density is $n_i=\Gamma n_i^*$ and pick-up of
interstellar protond and electrons occurs at a rate
\be
\dot {N}_i(\gamma )=\pi r^2cn_i^*\sqrt{\Gamma ^2-1}\delta (\gamma-\Gamma)
\label{pickuprate}
\ee
The pick-up is a source of isotropic, quasi-monoergetic protons and electrons with Lorentz
factor $\Gamma $ in the jet outflow frame. By interacting with the pair background
plasma (electron bremsstrahlung, proton inverse bremsstrahlung), the ordered magnetic field 
(synchrotron radiation) and internal and external target photon field 
(inverse Compton scattering) the pick-up electrons and protons generate nonthermal photons.
The detailed modelling of the radiation signature of this leptonic jet outflow,  
including the time evolution of the energy spectra of radiating particles and 
the transformation to the observer's frame, is analogous to the earlier treatment of 
the hadronic jet outflow (see Pohl \& Schlickeiser \cite{ps00}), but lies beyond the scope
of the present paper. We note, however, one important difference: the pick-up protons
do not find thermal protons in the background pair plasma to undergo 
inelastic $p-p$-collisions, so that no secondary neutral pions, secondary electrons and 
positrons and neutrinos will be produced. These radiation products, that provided the
dominating contribution of the Doppler-boosted GeV-TeV photon emission in the hadronic jet 
model and gave rise to correlated Doppler-boosted GeV-TeV neutrino emission 
(Schuster et al. \cite{sps02}), will be absent here. The pick-up protons, that carry a
factor $m_p/m_e=1836$ more power than the pick-up electrons, can only radiate 
by the much less efficient inverse bremsstrahlung process. High-energy GeV-TeV emission
in the leptonic jet outflow model will be dominated by inverse Compton scattering of 
pick-up electrons off internal and external target photon fields. It will be most 
interesting to study in future work quantitatively the leptonic jet radiation 
signatures modifying the formalism of B\"ottcher et al. (\cite{bsm01}) developed for 
optically-thick collision-dominated relativistic pair 
gamma-ray burst fireballs.
\subsection{Build-up of transverse magnetohydrodynamic turbulence}
The co-moving time scales 
\be
t_{t,e}=0.19
{(n^*_{b,10})^{1/2}\over n_i^*\Gamma _2^{1/6}}\hbox{   s}
\label{tpegrb}
\ee
and
\be
t_{t,p}=({m_p\over m_e})^{4/3}t_{t,e}=4.3\cdot 10^3
{(n^*_{b,10})^{1/2}\over n_i^*\Gamma _2^{1/6}}\hbox{   s}
\label{tppgrb}
\ee
given in Eqs. (\ref{tTe}) and 
(\ref{tTp}), respectively, also provide estimates for the time to build up
the fully developed power spectrum of transverse hydromagnetic turbulence 
from the beam electrons and the beam protons, respectively. Hydromagnetic 
turbulence upstream and downstream is crucial for diffusive shock acceleration
to operate. We recall that beam electrons alone only produce transverse
turbulence at wavenumbers larger than the inverse of the electron gyroradius
($|k|\ge R_e^{-1}$) with an (integrated) energy density smaller by a factor
$(m_e/m_p)=1/1836$ than the turbulence produced by the beam protons.
As our earlier estimate (\ref{coneiso}) indicated, this small 
fraction of electron 
generated transverse turbulence is built up after time $t_{t,e}$ for
standard AGN outflow and environment parameters. For the protons 
to provide the much larger fraction of transverse turbulence, the longer
time $t_{t,p}$ is needed, and it will occur at later times of the AGN jet
evolution. The co-moving time scales (\ref{tpegrb})-(\ref{tppgrb}) correspond to 
Doppler-shortened time scales $t_t^*=t_t/D_L$, where the Doppler factor 
$D_L=\Gamma ^{-1}[1-(V\cos \theta ^*/c)]^{-1}$, 
depends on the observer's viewing angle $\theta ^*$.

Eqs. (\ref{tpegrb}) and (\ref{tppgrb}) indicate that it takes
a non-negligible fraction of the early AGN jet phase,
to build up the magnetohydrodynamic turbulence for nonthermal particle
acceleration at either internal or external shocks associated with the pair jet outflow.

\section{Summary and conclusions}
We have investigated the
microphysical details of the energy conversion in relativistic pair outflows
interacting with the surrounding interstellar medium consisting of cold protons
and electrons. We have represented the relativistic pair blast wave 
as a one-dimensional channeled outflow directed parallel to a uniform 
background magnetic field. 
Viewed from the coordinate system
comoving with the pair outflow, the interstellar protons and electrons 
represent a
proton-electron beam propagating with relativistic speed  
antiparallel to the
uniform magnetic field direction. We demonstrate that the beam 
excites both electrostatic and 
low-frequency magnetohydrodynamic Alfven-type waves via a two-stream
instability in the pair background plasma, and we calculate the time evolution 
of the distribution functions of the
beam particles and the generated
plasma wave turbulence power spectra.

For standard AGN jet outflow and environment parameters we show that
the initial 
beam distributions of interstellar protons and electrons quickly relax
to  plateau-distributions 
in parallel momentum, transferring thereby one-half 
of the initial energy density of the beam particles to electric field
fluctuations of the generated electrostatic turbulence.

On considerably longer time scales,
the plateaued  interstellar electrons and protons will isotropise by their self-generated
transverse turbulence and thus be picked-up in the outflow pair plasma.
These longer time scale are also characteristic for establishing 
fully developed power spectra of transverse hydromagnetic turbulence 
from the plateaued electrons and protons. This hydromagnetic 
turbulence upstream and downstream is crucial for diffusive shock acceleration
to operate at external or internal shocks associated with pair outflows.  It 
takes a finite time period of the early AGN outflow phase 
to build up the magnetohydrodynamic turbulence for nonthermal particle
acceleration at either internal or external shocks associated with pair jet outflows.
During this initial phase -- prior to the generation of hydromagnetic
turbulence -- the radiative and energy-exchange processes in the jet outflow are not 
dominated by acceleration 
processes of nonthermal charged particles at collisionless shock waves. 

\begin{acknowledgements}
We gratefully acknowledge the help\-ful and constructive comments of the referee. 
RS and IL ack\-nowledge support by the
Deutsche For\-schungs\-ge\-mein\-schaft through Sonderforschungsbereich 191.
RV acknowledges the financial support of the Academy of Finland (project
46331) and the PLATON network (EC contract HPRN-CT-2000-00153).
The work of MB is supported by NASA through Chandra Postdoctoral 
Fellowship grant PF~9-10007 awarded by the Chandra X-ray Center, 
which is operated by the Smithsonian Astrophysical Observatory 
for NASA under contract NAS~8-39073. 
RS, CS and MP acknowledge support by the Verbundforschung, grants
DESY AG9PCA7 and DESY CH1PCA6.
\end{acknowledgements}

\section{Appendix A: Dispersion relations of parallel plasma waves in a 
thermal magnetised pair plasma}
The properties of parallel propagating plasma waves in magnetised thermal 
plasmas have been discussed by Fichtner \& Schlickeiser (\cite{fs95}) and
Schlickeiser et al. (\cite{sfk97}). For subluminal waves with index of
refraction $N=ck/\omega _R>1$ the dispersion relations of longitudinal ($\Lambda _L=0$) 
and transverse ($\Lambda _T=0$) can be split in their real and imaginary parts.

\subsection{Longitudinal waves}

According to Eq. (22) of Fichtner \& Schlickeiser (\cite{fs95}) we obtain for
longitudinal waves in the limit of weak damping
$\omega =\omega _R+\imath \omega _I $, $\omega _I<<\omega _R$, 
\bdm
\Re \Lambda _L(\omega _R,k)=
1+{\sigma ^2(1+2\sigma ^2)\over \omega _R^2}\sum _a\omega _{p,a}^2\mu _a\; -
\edm
\bdm 
\sum _a {\omega _{p,a}^2\mu _a^2\sigma ^3\over \omega _R^2K_2(\mu _a)}
\int_0^{\sigma }dy\; (1+\sigma ^2+y^2){K_3(\mu _a\sqrt{1+\sigma ^2-y^2}\over 
(1+\sigma ^2-y^2)^{3/2}}
\edm
\be
=0
\label{relong1}
\ee
and
\bdm 
\Im \Lambda _L(\omega _R,k)={\pi \over 2}\rm {sgn} (\omega _R)\sum _a
{\omega _{p,a}^2\exp (-\mu _a\sqrt{1+\sigma ^2})\over \omega _R^2N^3\mu _aK_2(\mu _a)}
\edm 
\be 
\bigl[2+2\mu _a\sqrt{1+\sigma ^2}+\mu _a^2(1+\sigma ^2)\bigr]=0
\label{imlong1}
\ee
where $\sigma =(N^2-1)^{-1/2}$,
$\mu _a=1/\Theta _a=m_ac^2/(k_BT_a)$, and $\omega _{p,a}$ denotes the plasma 
frequency of species $a$. Solutions of Eq. (\ref{relong1}) define the frequency-wavenumber 
relations $\omega _R=\omega _R(k)$ of all possible modes in the system whereas
the growth or damping rate of each mode is calculated from Eqs. (\ref{relong1}) and 
(\ref{imlong1}) by the standard weak-damping expansion
\be
\omega _I=-{\Im \Lambda _L(\omega _R,k)\over 
[{\partial \Re \Lambda _L(\omega _R,k)\over \partial \omega _R}]}
\label{omigl1}
\ee

The integral equation ((\ref{relong1}) has been further reduced by 
Schlickeiser \& Mause (\cite{sm95}). 
Introducing normalised wavenumbers $\kappa =ck/\omega _{p,e}$ and 
frequencies $f=\omega _R/\omega _{p,e}$ 
these authors obtained for  non-relativistic temperatures
$\mu _a>>1$ (see Eqs. (34)-(38) of Schlickeiser \& Mause (\cite{sm95}))
the approximation
\be
\Re \Lambda _L(f,\kappa )\simeq 
1-\sum _a{\xi _a\over f^2}\bigl[1-{5\over 2\mu _a}+{3\kappa ^2\over \mu _af^2}\bigr]
=0
\label{relong2}
\ee
with $\xi _a=\omega _{p,a}/\omega _{p,e}$. 
For a single temperature plasma this approximation holds for the wavenumber range
\be
[\sum _a(1+{1\over 2\mu _a})]^{1/2}\le |\kappa |\le 2\mu _a^{1/2}[1-{5\over 16\mu _a}]
\label{krangelon}
\ee

For a single-temperature ($\mu _a=1/\Theta $) non-relativistic electron-positron plasma to 
lowest order in the normalised temperature $\Theta <<1$, Eq. (\ref{relong2})
reduces to 
\be
\Re \Lambda _L(\omega _R,k)\simeq 1-{2\omega ^2_{p,e}\over \omega _R^2}=0
\label{relong3}
\ee
yielding electrostatic waves
\be
\omega _R^2=2\omega ^2_{p,e}
\label{eposdis1}
\ee
in the wavenumber range $|k|\le 2\omega _{p,e}/(c\Theta ^{1/2})$
as only longitudinal plasma mode. Moreover, we obtain
\be
{\partial \Re \Lambda _L(\omega _R,k)\over \partial \omega _R}=
{4\omega _{p,e}^2\over \omega _R^3}={2\over \omega _R}
\label{partabl}
\ee
While superluminal electrostatic waves are not damped, in the 
subluminal wavenumber range $\sqrt{2}\omega _{p,e}/c\le |k|\le 2\omega _{p,e}/(c\Theta ^{1/2})$
we derive for the associated damping rate (\ref{imlong1})
\bdm
\omega _I=-{\omega _R\over 2}\Im \Lambda _L(\omega _R,k)=
-\pi \rm {sgn} (\omega _R){\omega _{p,e}^4\Theta \over c^3k^3K_2(1/\Theta )}
\edm 
\bdm 
\exp (-\sqrt{1+\sigma ^2}/\Theta )
\bigl[2+{2\sqrt{1+\sigma ^2}\over \Theta }
+{(1+\sigma ^2)\over \Theta ^2}
\bigr] 
\edm
\bdm 
=-\pi \rm {sgn}(\omega _R) 
{\omega _{p,e}^4\over ck[c^2k^2-2\omega _{p,e}^2]\Theta K_2(1/\Theta )}
\edm 
\bdm 
\exp \Bigl[-{c|k|\over \Theta \sqrt{c^2k^2-2\omega _{p,e}^2}}\Bigr]
\edm
\be
\Bigl[1+\; {2\Theta \over c|k|}\sqrt{c^2k^2-2\omega _{p,e}^2}+\; 
{2\Theta ^2(c^2k^2-2\omega _{p,e}^2)\over c^2k^2}\Bigr]
\label{dampingelect1}
\ee
\subsection{Transverse waves}
The dispersion relation of parallel propagating transverse waves
in a thermal plasma with non-relativistic temperatures has been analysed in Sect. 3
of Schlickeiser et al. (\cite{sfk97}). Its real part
\be
\Re \Lambda _T=1-N^2-\sum _a{\omega _{p,a}^2\over kv_{th,a}\omega _R}
Z\Bigl({\omega _R-\Omega _a\over kv_{th,a}}\bigl)=0
\label{retrans1}
\ee
involves the plasma dispersion function $Z(x)$ (Fried \& Conte \cite{fc61}), the
non-relativistic gyrofrequencies  
$\Omega _a=q_aB/(m_ac)$ and the non-relativistic thermal velocities 
$v^2_{th,a}=2c^2/\mu _a=2k_BT_a/m_a$.
In Eq. (\ref{retrans1}) we adopt the convention 
that positive values of $\omega _R>0$ denote left-handed circularly polarized 
waves, and that negative values of $\omega _R<0 $ denote right-handed 
circularly polarized waves.
An expansion of $Z(x)$ for large arguments and the weak-damping approximation 
$|\omega _I/\omega _R|<<1$ allows the derivation of a simplified, asymptotic dispersion relation
for transverse oscillations propagating parallel to the external magnetic field
\be
\Re \Lambda _T\simeq 1-{c^2k^2\over \omega _R^2}-\; 
\sum _a{\omega _{p,a}^2\over 
\omega _R(\omega _R-\Omega _a)}=0
\label{retrans2}
\ee
According to Eq. (54) of Schlickeiser et al. (\cite{sfk97}) we obtain for
the weak damping rate in the same limit
\bdm 
\omega _I=
-{\pi ^{1/2}\over \omega _R}\rm {sgn} (\omega _R)
\Bigl[{\partial \Re \Lambda _T(\omega _R,k)\over \partial \omega _R}\Bigr]^{-1}
\edm 
\be 
\sum _a
{\omega _{p,a}^2\over kv_{th,a}}
\exp \bigl(-[{\omega _R-\Omega _a\over kv_{th,a}}]^2\bigr)
\label{dampingtrans1}
\ee
For a single-temperature ($\mu _a=1/\Theta $) non-relativistic electron-positron plasma 
Eq. (\ref{retrans2})
reduces to 
\be
\Re \Lambda _T\simeq 1-{c^2k^2\over \omega _R^2}-\; 
{2\omega _{p,e}^2\over 
\omega ^2_R-\Omega _e^2}=0
\label{retrans3}
\ee
For subluminal phase speeds ($|\omega _R/k|<<c$) we may neglect the first term
on the right hand side of Eq. (\ref{retrans3}), so that the
transverse dispersion relation yields 
\be 
\omega _R^2\simeq {V_e^2\Omega _e^2k^2\over \Omega _e^2+V_e^2k^2}
\label{disprt1}
\ee
where
\bdm 
V_e={\Omega _ec\over \sqrt{2}\omega _{p,e}}=
B\;[8\pi n_bm_e]^{-1/2}=
\edm 
\bdm 
6.6\cdot 10^{12}(B/\rm {Gauss} )(n_b/\rm {cm} ^{-3})^{-1/2}
\; {\rm cm}\; {\rm s}^{-1}=
\edm
\be
9.2\cdot 10^7(\epsilon_{B,-1}\Theta _{-4})^{1/2}
\; {\rm cm}\; {\rm s}^{-1}
\label{alf1}
\ee
is the Alfven speed in the pair plasma. For wavenumbers much less than 
$|k|<<k_e$, where 
\be
k_e\equiv \Omega _e/V_e
=2.7\cdot 10^{-2}n_{b,8}^{1/2}
\; {\rm cm}^{-1}
\label{pairskin1}
\ee
is the inverse pair skin length, Eq. (\ref{disprt1}) reduces to 
\be 
\omega _R^2=V_e^2k^2
\label{alfwave}
\ee
whereas in the opposite case $|k|>>k_e$ Eq. (\ref{disprt1}) 
approaches the pair-cyclotron wave limit $\omega _R^2\simeq \Omega _e^2$.

Moreover, Eq. (\ref{retrans3}) yields
\be
{\partial \Re \Lambda _T(\omega _R,k)\over \partial \omega _R}=
{4\omega _{p,e}^2\omega _R\over (\omega _R^2-\Omega _e^2)^2}
+\; {2c^2k^2\over \omega _R^3}
\label{partablt1}
\ee
so that the damping rate (\ref{dampingtrans1}) becomes
\bdm 
\omega _I(k)=
-{1\over 2}\sqrt{\pi \over 2\Theta }
{\omega ^2_{p,e}\over ck}
{\rm {sgn} (\omega _R)\over 
[{c^2k^2\over \omega _R^2}+{2\omega _{p,e}^2\omega _R^2\over (\omega _R^2-\Omega _e^2)^2}]}
\edm 
\be
\Bigl[\exp \bigl(-{(\omega _R-\Omega _e)^2\over 2\Theta k^2c^2}\bigr)
+\; \exp \bigl(-{(\omega _R+\Omega _e)^2\over 2\Theta k^2c^2}\bigr)\Bigr]
\label{dampingtrans2}
\ee
Especially, for small frequencies ($|\omega _R|<<\Omega _e$) we obtain for the 
pair Alfven waves
(\ref{alfwave}) the damping rate
\bdm
\omega _{I,A}(k)= 
-{\pi ^{1/2}\over 2}{\Omega _e^2\over v_{th,e}|k|(1+{k^2\over k_e^2})}
\exp \bigl(-({\Omega _e\over v_{th,e}k})^2\bigr)
\edm
\be
\simeq 
-{\pi ^{1/2}\over 2}\; {\Omega _eV_e\over v_{th,e}}\; 
{k_e\over |k|}
\exp \bigl(-({V_ek_e\over v_{th,e}k})^2\bigr)
\label{dampingalfv}
\ee
where $v_{th,e}=\sqrt{2\Theta }c=\sqrt{2k_BT_e/m_e}$ is the non-relativistic 
thermal pair velocity. 

To check the weak-damping approximation we calculate  
\be
|\omega _R/\omega _I|={\pi ^{1/2}\over 2}\beta ^{1/2}_p{k_e^2\over k^2}
\exp [-{\beta _pk_e^2\over k^2}]
\label{checkweakdamping}
\ee 
where we introduce the pair plasma beta $\beta _p=V_e^2/v_{th,e}^2=0.05\epsilon _{B,-1}$.
One notices that for small wavenumber values 
$|k|<<\beta _p^{1/2}k_e=0.22\epsilon _{B,-1}^{1/2}k_e$
the weak-damping approximation is well satisfied.
\section{Appendix B: Comparison of the initial growth rates with the thermal 
Landau and cyclotron damping rates}
We noted that plasma waves, due to the unstable interstellar beam particles entering the
jet outflow pair plasma will only grow if the initial growth rate is larger than the 
thermal Landau (\ref{dampingelect1}) and cyclotron (\ref{dampingalfv}) damping rates.
Here we investigate under which conditions this is the case. We consider longitudinal and
transverse waves in turn.
\subsection{Electrostatic waves}
If we include the contributions from the beam particles the initial longitudinal 
dispersion relation reads
\bdm 
\Lambda _e(t=0)=\Re \Lambda _L+\; \imath \Im \Lambda _L+
\; {2\pi \over \omega }\sum _i\omega _{p,i}^2
\edm
\be 
\times \int _{-\infty }^\infty dp_{\parallel }
\int _0^\infty dp_{\perp }\; {p_{\perp }p_{\parallel }\over \gamma (\omega -kv_{\parallel })}
\; {\partial f_i(p_{\perp }, p_{\parallel},t=0)\over \partial p_{\parallel}}
\label{totdispe}
\ee
where $\Re \Lambda _L$ and $\Im \Lambda _L$ are given by Eqs. (\ref{relong3}) and 
(\ref{imlong1}), respectively. Neglecting $\Im \Lambda _L$ for this calculation 
we obtain in terms of the normalised phase space distribution (\ref{distzyl})
\bdm
\Lambda _e(t=0)\simeq \Re \Lambda _L+\; 
{1\over \omega }\sum _i\omega _{p,i}^2
\int _{-\infty }^\infty dp_{\parallel }\; p_{\parallel }
\edm
\be
\bigl[(\sqrt{1+{p^2_{\parallel} \over m_i^2c^2}}
\omega -{kp_{\parallel }\over m_i})\bigr]^{-1}
\; {\partial G_i(p_{\parallel},t=0)\over \partial p_{\parallel}}
\label{totdispe1}
\ee
Inserting the initial distribution function (\ref{inif}), i.e. 
$G(p_{\parallel},t=0)=\delta (p_{\parallel}+P)$, and integrating partially with respect to 
$p_{\parallel }$ we derive with Eq. (\ref{relong3})
\be
\Lambda _e(t=0)=1-\; {2\omega _{p,e}^2\over \omega ^2}-
\; {\sum _i\omega _{p,i}^2\over \Gamma ^3(\omega +kV)^2}=0
\label{totdispe2}
\ee
With the abbreviations
\be
w_e=\sqrt{2}\omega _{p,e},\;\;\; w_i=(\sum _i\omega _{p,i}^2)^{1/2}
\label{abbe1}
\ee
and the substitution 
\be
f=\omega +kV
\label{fdef1}
\ee
Eq. (\ref{totdispe2}) becomes
\be
|kV-f|={w_e\over (1-{w_i^2\over \Gamma ^3f^2})^{1/2}}
\label{totdispe3}
\ee
A posteriori it can be shown that in the resonant wave regime
\be
w_i^2<<\Gamma ^3f^2
\label{resoregime}
\ee
so that the sqare root in Eq. (\ref{totdispe3}) can be expanded:
\be
kV\simeq f-\; w_e-\; {w_ew_i^2\over 2\Gamma ^3f^2}
\label{bune1}
\ee
This equation has already been examined by Buneman (\cite{b59}).
Substituting
\be
f={w_e^{1/3}w_i^{2/3}\over \Gamma }x,\;\;\;
kV=-w_e+{w_e^{1/3}w_i^{2/3}y\over \Gamma }
\label{subst1}
\ee
Eq. (\ref{bune1}) reduces to
\be
x^3-\; yx^2-\; {1/2}=0
\label{cubic1}
\ee
This cubic equation has one real and two complex conjugate solutions $x_{1,2}$ provided 
$y>-3/2$, corresponding to wavenumbers
\be
k>-[{3w_e^{1/3}w_i^{2/3}\over 2\Gamma V}+{w_e\over V}]
\label{klrang1}
\ee
The complex conjugated solutions are
\bdm
x_{1,2}={y\over 3}-\; {1\over 2}\Bigl[({1\over 4}+{y^3\over 27}+\Delta ^{1/2})^{1/3}+
({1\over 4}+{y^3\over 27} -\Delta ^{1/2})^{1/3}\Bigr]
\edm
\be
\pm \imath {\sqrt{3}\over 2}
\Bigl[({1\over 4}+{y^3\over 27} +\Delta ^{1/2})^{1/3}-
({1\over 4}+{y^3\over 27} -\Delta ^{1/2})^{1/3}\Bigr]
\label{complsol1}
\ee
where
\be
\Delta ={1\over 16}+{y^3\over 54}
\label{Deltavalue}
\ee
The maximum growth rate
$\Im x_{max}=3^{1/2}2^{-4/3}$
occurs for $y=0$, corresponding to 
\be
k_0=-w_e/V=-2^{1/2}\omega _{p,e}/V
\label{k0def}
\ee
Consequently 
\bdm
\omega _{I,max}=\Im \omega _{max}=\Im f_{max}=
\edm 
\bdm 
{w_e^{1/3}w_i^{2/3}\over \Gamma }\Im x_{max}={3^{1/2}\over 2\Gamma }
\omega _{p,e}^{1/3}(\sum _i\omega _{p,i}^2)^{1/3}=
\edm
\be
{3^{1/2}\over 2}{\omega _{p,e}\over \Gamma}[{n_i\over n_b}(1+{m_e\over m_p})]^{1/3}
={3^{1/2}\over 2}\omega _{p,e}[{n^*_i\over \Gamma n^*_b}]^{1/3}
\label{maxgrow}
\ee
Now we compare this maximum growth rate with the Landau damping rate (\ref{dampingelect1})
calculated at the wavenumber $k_0$ which is
\bdm
\omega _I(k_0)=-{\pi \over 2^{3/2}}\omega _{p,e}{(\Gamma ^2-1)^{3/2}\over \Gamma }
{e^{-\Gamma /\Theta }\over \Theta K_2(1/\Theta )}
[1+
\edm
\be 
{2\Theta \over \Gamma }+{2\Theta ^2\over \Gamma^2}]\simeq 
-{\pi ^{1/2}\over 2\Theta ^{3/2}}\omega _{p,e}\Gamma ^2\exp [-{\Gamma -1\over \Theta }]
\label{landauatk0}
\ee
For the ratio of the maximum growth rate to this damping rate we obtain
\bdm
{\omega _{I,max}\over |\omega _i(k_0)|}=
\sqrt{3/\pi }({n_i^*\over n_{b,10}^*})^{1/3}\Theta ^{3/2}\Gamma ^{-7/3}
\exp (-{\Gamma -1\over \Theta })=
\edm
\be
(n_i^*)^{1/3}(n_{b,10}^*)^{1/3}\Theta _{-4}^{3/2}\Gamma _2^{-7/3}
10^{4.23\cdot 10^5\Gamma _2\Theta _{-4}^{-1}-18}>>1
\label{ratiodamplan1}
\ee
which is a huge factor for standard AGN jet outflow and environment parameters due to 
the exponential dependence on $\exp [(\Gamma-1)/\Theta ]$.
So we can safely neglect the Landau damping of the beam generated electrostatic waves.
Only for values of $\Gamma _2<4\cdot 10^{-5}\Theta _{-4}$, i.e. very hot
pair plasmas and slow Lorentz factors of the outflow, the damping of electrostatic waves
outnumbers the initial wave growth.
\subsection{Transverse waves}
Including the contributions from the beam particles to the initial transverse 
dispersion relation we obtain after analogous manipulations as in the previous subsection 
\bdm 
\Lambda _t(t=0)=1-\; {c^2k^2\over \omega ^2}-\; {2\omega _{p,e}^2\over \omega ^2-\Omega _e^2}-
\; {\omega _{p,e}^2n_i(\omega +kV)\over \omega ^2\Gamma n_b}
\edm
\be
\bigl[
{m_e\over m_p(\omega +kV-{m_e\Omega _e\over m_p\Gamma })}+
{1\over \omega +kV+{\Omega _e\over \Gamma }}\bigr]=0
\label{totdispt1}
\ee
where the two beam contributions stem from the interstellar protons and electrons, respectively.
The beam protons will excite transverse waves near their resonance
\be
\omega _R+kV-{m_e\Omega _e\over m_p\Gamma }=0
\label{protonres}
\ee
whereas the beam electrons excite transverse waves near 
\be
\omega _R+kV+{\Omega _e\over \Gamma }=0
\label{elecres}
\ee
Here we consider only the waves generated by the interstellar protons, and leave the case of
electron generated waves to the interested reader. We also assume small wave frequencies,
i.e. $\omega ^2<<\Omega _e^2$ so that Eq. (\ref{totdispt1}) reduces to 
\bdm 
\Lambda _t(t=0)\simeq 1-\; {c^2k^2\over \omega ^2}+\; {c^2\over V_e^2}
\edm 
\be 
-\; {\omega _{p,e}^2n_i\over \omega ^2\Gamma n_b}
{\omega +kV\over \omega +kV-{m_e\Omega _e\over m_p\Gamma }}=0
\label{totdisptp1}
\ee
This approximation holds for wavenumbers $k$ near the proton resonace condition (\ref{protonres}).
Using $V_e^2<<c^2$ we then obtain
\be
0=\omega ^2-\; V_e^2k^2-\; 
\alpha \omega _{p,e}^2{\omega +kV\over \omega +kV-{m_e\Omega _e\over m_p\Gamma }}
\label{pdisp1}
\ee
where we introduce the parameter
\be
\alpha ={V_e^2n_i\over c^2\Gamma n_b}
\label{alfadef1}
\ee
Eq. (\ref{pdisp1}) leads to the cubic equation
\be
\omega ^3+b\omega ^2+g\omega +h=0
\label{cubict}
\ee
with
\bdm 
b=kV-\Omega ,\;\;\; g=-(V_e^2k^2+\alpha \omega _{p,e}^2),
\edm
\be
h=kVg+\Omega V_e^2k^2,\;\;\; \Omega ={m_e\Omega _e\over m_p\Gamma }
\label{cubict1}
\ee
Provided
\be
D=Q^2+S^3>0,
\label{Deq1}
\ee
where
\bdm 
Q=(b/3)^3-(bg/6)+(h/2)=
\edm 
\be 
{1\over 27}(kV-\Omega )^3+{2\over 3}kVg-\Omega g+{1\over 2}\Omega V_e^2k^2
\label{qtdef}
\ee
and
\be
S=(g/3)-(b/3)^2=-{1\over 9}[(kV-\Omega )^2+3g],
\label{ptdef}
\ee
the cubic equation (\ref{cubict}) has two complex conjugate solution 
\bdm 
\omega ={1\over 3}(\Omega -kV)-{1\over 2}[(-Q+\sqrt{D})^{1/3}+(-Q-\sqrt{D})^{1/3}]
\edm 
\be 
\pm {\imath \sqrt{3}\over 2}[(-Q+\sqrt{D})^{1/3}-(-Q-\sqrt{D})^{1/3}]
\label{cubictsol}
\ee
The maximum growth rate is a measure for the importance of the instability. For its estimate we
use the ansatz of Achatz et al. (\cite{als90}) that it is 
largest when the resonance condition (\ref{protonres}) 
is fulfilled yielding with $\omega _R=\pm V_ek$ for relativistic outflow velocities ($V>>V_e$)
\be 
k=k_0={\Omega \over V\pm V_e}\simeq {\Omega \over V}
\label{k0tdef}
\ee
For this wavenumber we obtain
\be
S(k_0)={1\over 3}\alpha \omega _{p,e}^2[1+3r],\;\;\;
Q(k_0)={\Omega \over 3}\alpha \omega _{p,e}^2[1+{5r\over 2}]
\label{pk0}
\ee
with the frequency ratio
\bdm 
r={V_e^2\Omega ^2\over \alpha \omega _{p,e}^2V^2}=
{2m_e^2\over m_p^2}{V_e^2\over c^2\Gamma (\Gamma ^2-1)}{n_b^*\over n_i^*}
\edm 
\be 
=5.6\cdot 10^{-8}{n_{b,10}^*\over n_i^*}\epsilon _{B,-1}\Theta _{-4}\Gamma _2^{-3}
\label{freqratt}
\ee
which for standard jet outflow and environment parameters is very small.
We then obtain
\bdm
\omega _I(k_0)=\Im \omega (k_0)={3^{1/2}\over 2}D^{1/6}(k_0)
\edm 
\bdm 
\times \bigl[(1+{Q(k_0)\over D^{1/2}(k_0)})^{1/3}+
(1-{Q(k_0)\over D^{1/2}(k_0)})^{1/3}\bigr]
\edm 
\be 
\simeq 3^{1/2}D^{1/6}(k_0)\simeq 
(\alpha /3)^{1/2}[1+{3\Omega ^2\over \alpha \omega _{p,e}^2}]^{1/6}
\label{maxtgrow}
\ee
because
\be
{Q^2(k_0)\over D(k_0)}={r\over r+(V_e/V)^2}<1
\label{begr1}
\ee
is always smaller than unity.

Collecting terms we obtain
\bdm 
\omega _I(k_0)={\omega _{p,e}\over \sqrt{3}}{V_e\over c}
({n_i^*\Gamma \over n_b^*})^{1/2}[1+0.018{n_{b,10}^*\Gamma _2^3\over n_i^*}]
\edm 
\be 
=10^{-4}{\Omega _e\over \sqrt{6}}
({n_i^*\Gamma _2\over n_{b,10}^*})^{1/2}[1+0.018{n_{b,10}^*\Gamma _2^3\over n_i^*}]
\label{growt2}
\ee
which has to be compared with the cyclotron damping rate (\ref{dampingalfv})
calculated at the wavenumber $k_0$:
\be
\omega _{I,A}(k_0)=-{\pi ^{1/2}\over 2}\Omega _e\; d_1e^{-d_1^2}
\label{alfvtd1}
\ee
where
$d_1=1.3\cdot 10^7\Gamma _2\Theta {-4}^{-1/2}$. Because of the small factor
$d_1\exp (-d_1^2)=\exp (-1.7\cdot 10^{14})$ the cyclotron damping rtae is negligibly
small compared to the initial growth rate.

\section{Appendix C: Calculation of the wave spectrum with the bi-spherical
shell method}
Here the self-generated wave spectrum of backward moving transverse waves 
can be
evaluated with the assumption, that the free energy (as measured in
the plasma rest frame) of the initial beam particle distribution is
given to these waves in the isotropization process. This method has
been used before for non-relativistic particles when evaluating the
wave spectrum generated by freshly ionized pick-up ions (e.g.,
Huddleston \& Johnstone \cite{Hudd92}; Isenberg \& Lee \cite{Isen96}).
Generalizing this method to relativistic particles and waves with
non-relativistic phase speeds goes straight-forwardly by replacing
particle speeds with their momenta.

The final particle distribution is uniform in the so-called
(Huddleston \& Johnstone \cite{Hudd92}) bi-spherical shell, which in
our case of almost Alfv\'enic waves and particles injected at
$p_\perp=0$ is accurately given by the spherical shell
\begin{equation}
p_\perp^2+p_\parallel^2=p^2,
\end{equation}
with the momentum measured in the Alfv\'en wave frame, where its
magnitude, $p$, is taken as constant. (The final distribution function
is, thus, isotropic in the Alfv\'en-wave frame.) We may write the
self-generated wave spectrum in form (Isenberg \& Lee \cite{Isen96})
\begin{equation}
I_\alpha(k) = 4\pi \frac{dE}{dk} N_ \alpha\{p_\parallel(k)\},
\end{equation}
where $dE/dk = |dE/dp_\parallel||dp_\parallel/dk|$ is the energy lost
by a particle (per unit resonant wavenumber) as it scatters through
the parallel momentum range resonant with wavenumber range $(k,k+dk)$,
and
\begin{equation}
N_\alpha \{p_\parallel(k)\} = n_\alpha \frac{p-p_\parallel(k)}{2p},
\end{equation}
is the number density of particles scattered accross the parallel
momentum $p_\parallel(k)$, resonant with $k$, during the isotropization.
We have $|dE/dp_\parallel|=V_e$ (with $E$ still measured in the plasma
frame, of course), so
\begin{equation}
I_\alpha (k) = 2\pi n_\alpha V_e\left|\frac{dp_\parallel}{dk}\right|\frac{p-p_\parallel}{p},
\end{equation}
where one uses Eq.(\ref{mures}) (and $p_{\parallel }=p\mu _i$) 
evaluated in the Alfv\'en-wave
frame to get
\begin{equation}
p_\parallel(k)\approx-\frac{m_\alpha\Omega_\alpha}{k}+\gamma\,m_\alpha V_e\frac{k^2}{2k_e^2}
\end{equation}
for the $k$-range of interest to our analysis. Thus, the spectrum reads
\bdm
I_\alpha(k) = 2\pi n_\alpha p V_e\bigl(\frac{1}{R_{L,\alpha} k^2}
\edm 
\be 
+\epsilon_\alpha\frac{V_e}{v}\frac{k}{k_e^2}\bigr)
\bigl(1 +\frac{\epsilon_\alpha}{R_{L,\alpha}k}\bigr),
\ee 
where $R_{L,\alpha}=\gamma v/|\Omega_\alpha|$ is the particle's Larmor
radius and $\epsilon_\alpha=\mathrm{sign}(\Omega_\alpha)$. The small
correction due to the dispersive effects in $dp_\parallel/dk$ has to
be retained, but the corresponding term in $p-p_\parallel$ can be
neglected, since it is small in the region $p_\parallel\approx p$.

\begin{figure}
{\centering \resizebox*{1\columnwidth}{!}{\includegraphics{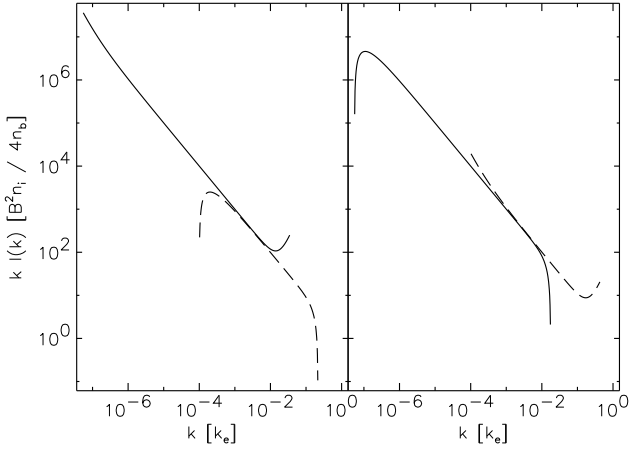}}}
\caption{Spectrum of self-generated waves for after a relativistic
pick up of electrons and protons with a $\Gamma=100$. The left [right]
panel gives the spectrum of the right-handed [left-handed]
waves. Electron-generated (proton-generated) waves are plotted with a
{\em dashed curve} ({\em solid curve}).}
\label{fig:I(k)}
\end{figure}

The above spectrum holds only for a single particle species being
picked up by the waves. When several particle species are picked up
simultanously, then it may not be possible to avoid considering
multiple resonances.  For the least massive species the reasoning
behind the neglecting of them is still valid, but for the more massive
ones, part of the free energy of the lightest species will go to the
``denied'' bands of the massive particles, which may invalidate the
assumption of single resonant wavenumber. In the specific case of
electrons and protons, however, the region where multiple resonance
are possible for protons, the electron-generated spectrum has a steep
slope ($\propto k^{-2}$), which as a first approximation still enables
us to neglect multiple resonances in case of protons, as well. Thus,
the total wave spectrum comes from a contribution from protons and
electrons as
\bdm 
I(k) = 2\pi\sum_\alpha H(|k|R_{L,\alpha}-1)\,n_\alpha p_\alpha V_e
\bigl(\frac{1}{R_{L,\alpha}k^2}
\edm 
\be
+\epsilon_\alpha\frac{V_e}{v_\alpha}\frac{k}{k_e^2}\bigr)
\bigl(1 +\frac{\epsilon_\alpha}{R_{L,\alpha}k}\bigr).
\ee 
Taking $n_\alpha=n_i$, $v_\alpha = V$, $p_\alpha=\Gamma m_\alpha V$
with $\Gamma = 1/\sqrt{1-(V/c)^2}$, and $R_{L,\alpha}=\Gamma
V/|\Omega_\alpha|$, gives
\bdm 
I(k) = \frac{B^2}{4k_e}\frac{n_i}{n_b} \sum_\alpha H(\Gamma V |k|-\Omega_\alpha) 
\edm 
\be 
\left(\frac{k_e^2}{k^2}+\epsilon_\alpha\frac{\Gamma m_\alpha}{m_e}\frac{k}{k_e}\right)\left(1 +\frac{\Omega_\alpha}{\Gamma V k}\right)
\ee 
showing that the electronic contribution at $|\Omega_e| / \Gamma
V|k|\ll(m_e/\Gamma m_p)^{1/3}k_e\ll k_e/\Gamma^{1/3}$ is of
the same order as the ionic one. The spectrum and its constituents are
plotted in Fig.~\ref{fig:I(k)}.

\section{Appendix D: Isotropisation by Coulomb interactions}
Apart from scattering with transverse plasma waves additional pitch angle scattering of
the plateaued interstellar electrons and protons 
is provided by the elastic Coulomb interactions with the pair electrons and positrons in the 
jet outflow plasma. Here we demonstrate that the Coulomb isotropisation time scales of 
the beam electrons 
and protons are much longer than the plasma wave isotropisation time scales (\ref{tTp}) and 
(\ref{tTe}). 

According to Kulsrud et al. (\cite{k73}) the linearized  
Fokker-Planck scattering operator for Coulomb interactions is
\be
\bigl({\partial f_i\over \partial t}\bigr)_C=
\nu _{ei}{\partial \over \partial \mu }(1-\mu ^2){\partial f_i
\over \partial \mu }
\label{coulombfp}
\ee
where the electron scattering rate due to 
M{\o}ller and Bhabha scattering (B\"ottcher et al. \cite{bsm01}) is
\be
\nu _{ee}=2.6\cdot 10^{-3}\; n^*_{b,10}\Gamma _2^{-1}\hbox{    Hz}
\label{coufre}
\ee
while the proton scattering rate (Mannheim \& Schlickeiser \cite{ms94}) is 
\be
\nu _{ep}=3.3\cdot 10^{-8}\; n^*_{b,10}\Gamma _2^{-1}\hbox{    Hz}
\label{coufrp}
\ee
The solution of the associated
Fokker-Planck equation during the secular phase
\be
\bigl({\partial f_i\over \partial t}\bigr)=
\bigl({\partial f_i\over \partial t}\bigr)_C
\label{fpcou}
\ee
with the initial condition (\ref{secini2}),
\be
f_i(p,\mu ,t_e)={1\over 2\pi Pp^2}\delta (\mu +1)H[P-p]
\label{secini3}
\ee
is easily obtained in terms of Legrendre polynomials as
\bdm 
f_i(p,\mu ,t\ge t_e)=
{H[P-p]\over 4\pi Pp^2}
\edm 
\be 
\times \sum_{n=0}^\infty (-1)^n(2n+1)P_n(\mu )\exp [-n(n+1)\nu _{ei}(t-t_e)]
\label{solfpcou}
\ee
For times larger than 
\be
t_{C,i}=(2\nu _{ei})^{-1}
\label{coutime}
\ee
Eq. (\ref{solfpcou}) approaches the isotropic distribution (\ref{secf})
\be
f_{i}(p)={H[P-p]\over 4\pi Pp^2}
\label{secf1}
\ee
here without generating transverse waves.

For electrons the Coulomb isotropisation time scale is
\be
t_{C,e}=
1.9\cdot 10^2\; \Gamma _2\; (n^*_{b,10})^{-1}
\hbox{    s}
\label{tcoue}
\ee
and for protons
\be
t_{C,p}=
1.5\cdot 10^7\; \Gamma _2\; (n^*_{b,10})^{-1}
\hbox{    s}
\label{tcoup}
\ee
We note that for both particles these time scales are much 
longer than the time scales
(\ref{tTe}) and (\ref{tTp}) for isotropisation by transverse plasma waves calculated 
for standard AGN jet outflow and environment parameter
\be
t_{t,e}=3.0\cdot 10^{-4}
\; {(n^*_{b,10})^{1/2}\over n_i^*\Gamma _2^{1/6}}\; \; \; \hbox{  s},
\label{tTeagn}
\ee
and
\be
t_{t,p}=(m_p/m_e)^{4/3}t_{t,e}=
6.7\; {(n^*_{b,10})^{1/2}\over n_i^*\Gamma _2^{1/6}}\; \; \; \hbox{  s}
\label{tTpagn}
\ee
{}

\end{document}